\documentclass{emulateapj}

\usepackage{amsmath} %for numbering appendix table correctly

\slugcomment{\textbf{Astronomical Journal, published}}

\newcommand{\sm}{M$_{\odot}$}

\newcommand{\ergscmA}{$~{\rm erg~s^{\scriptscriptstyle -1}~cm^{\scriptscriptstyle -1}~\AA^{\scriptscriptstyle -1}}$}

\newcommand{\kms}{\ensuremath{{\rm km~s}^{-1}}}

\newcommand{\snaj}{XRF~060218/SN~2006aj}

\newcommand{\He}{\ion{He}{1}}
\newcommand{\HeFour}{\ion{He}{1}~$\lambda$4471}
\newcommand{\HeFive}{\ion{He}{1}~$\lambda$5876}
\newcommand{\HeSix}{\ion{He}{1}~$\lambda$6678}
\newcommand{\HeSeven}{\ion{He}{1}~$\lambda$7065}

\newcommand{\HeOnemicron}{\ion{He}{1}~$\lambda$10830}
\newcommand{\HeTwomicron}{\ion{He}{1}~$\lambda$20581}

\newcommand{\SiSix}{\ion{Si}{2}~$\lambda$6355}

\newcommand{\synNi}{$^{56}$\rm{Ni}}

\newcommand{\nsntot}{73} 
\newcommand{\nsnspec}{73}

\newcommand{\nsnnew}{53} 

\newcommand{\nsnIIb}{13} 
\newcommand{\nsnIb}{26} 
\newcommand{\nsnIc}{19} 
\newcommand{\nsnIcbl}{11}
\newcommand{\nsnnotsure}{4}

\newcommand{\ntotalspec}{645} 
\newcommand{\avespecforallsn}{9}
\newcommand{\ntotalspecfast}{615}%

\newcommand{\nsnvmax}{44} 
\newcommand{\nsnvnomax}{29} 
\newcommand{\nsnrmax}{6}
\newcommand{\nspecsnvmax}{508} 
\newcommand{\avespecforsnvmax}{12} 

\newcommand{\snzmean}{0.0141} 
\newcommand{\snzstdev}{0.0097}

\begin{document}

\title{Optical Spectra of \nsntot\ Stripped-Envelope Core-Collapse Supernovae }

\author{M.~Modjaz\altaffilmark{1}, %
S.~Blondin\altaffilmark{2}, %
R.~P.~Kirshner\altaffilmark{3}, 
T.~Matheson\altaffilmark{4},
P.~Berlind\altaffilmark{5},
F.~B.~Bianco\altaffilmark{1}
M.~L. Calkins\altaffilmark{5}
P.~Challis\altaffilmark{3}, 
P.~Garnavich\altaffilmark{6},
M.~Hicken\altaffilmark{3},
S.~Jha\altaffilmark{7},
Y.~Q. Liu\altaffilmark{1},
G.~H.~Marion\altaffilmark{3}
}

\altaffiltext{1}{Center for Cosmology and Particle Physics, New York University, 4 Washington Place, New York, NY 10003, USA; mmodjaz@nyu.edu.}
 \altaffiltext{2}{Aix Marseille UniversitŽ, CNRS, LAM (Laboratoire d'Astrophysique de Marseille) UMR 7326, 13388, Marseille, France.}
 \altaffiltext{3}{Harvard-Smithsonian Center for Astrophysics, 60 Garden Street, Cambridge, MA, 02138, USA.}
 \altaffiltext{4}{National Optical Astronomy Observatory, 950 North Cherry Avenue, Tucson, AZ 85719, USA.}
 \altaffiltext{5}{F. L. Whipple Observatory, 670 Mt. Hopkins Road, P.O. Box 97, Amado, AZ 85645, USA.}
 \altaffiltext{6}{Department of Physics, 225 Nieuwland Science Hall, University of Notre Dame, Notre Dame, IN 46556}
 \altaffiltext{7}{Department of Physics and Astronomy, Rutgers University, 136 Frelinghuysen Road, Piscataway, NJ 08854, USA.}

\begin{abstract}
We present \ntotalspec\ optical spectra of \nsntot\ supernovae (SN) of Types IIb, Ib, Ic, and broad-lined Ic. All of these types are attributed to the core collapse of massive stars, with varying degrees of intact  H and He envelopes before explosion. The SN in our sample have a mean redshift  $<cz>$ = 4200 \kms. Most of these spectra were gathered at the Harvard-Smithsonian Center for Astrophysics (CfA) between 2004 and 2009. For \nsnnew\ SN, these are the first published spectra. The data coverage range from mere identification (1-3
spectra) for a few SN to extensive series of observations (10$-$30
spectra) that trace the spectral evolution for others, with an average of \avespecforallsn\ spectra per SN. For \nsnvmax\ SN of the \nsntot\ SN presented here, we have well-determined dates of maximum light to determine the phase of each spectrum. Our sample constitutes the most extensive spectral library of stripped-envelope SN to date. We provide very early coverage (as early as 30 days before $V$-band max) for photospheric spectra, as well as late-time nebular coverage when the innermost regions of the SN are visible (as late as 2 years after explosion, while for SN~1993J, we have data as late as 11.6 years). This data set has homogeneous observations and reductions that allow us to study the spectroscopic diversity of these classes of stripped SNe and to compare these to SN associated with gamma-ray bursts (GRBs).  We undertake these matters in follow-up papers.
\end{abstract}

\keywords{supernovae: general\\ }

\section{Introduction}\label{ch4_intro_sec}

Stripped-envelope core-collapse supernovae (SN) mark the diverse deaths of massive stars and are sometimes accompanied by long-duration Gamma-Ray Bursts (GRBs), while their nucleosynthesis products contribute to the universe's chemical enrichment. %and their SN remnants are detected over the full wavelength regime (e.g., Cas A, \citealt{kirshner77,rest08,krause08}). 
By stripped-envelope core-collapse SN \citep{clocchiatti97} (SESN), for short "stripped SN", we refer to SN of
Types IIb, Ib, Ic and broad-lined Ic (SN IIb, Ib, Ic and Ic-bl; reviews on spectral SN classification
and the historical background can be found in \citealt{filippenko97_review} and \citealt{matheson01}). In this empirical classification scheme, SN I do not show evidence of H lines,  SN Ib show
conspicuous lines of \ion{He}{1}, while SN Ic do not, and neither
subtype shows strong \ion{Si}{2}, nor  \ion{S}{2} absorption lines, the hallmark of SN Ia. SN IIb are transition
types, for which the Balmer lines decrease in strength over time,
while \He\ lines, characteristic of SN Ib, appear in their spectra
(e.g., \citealt{filippenko93}).

Now, stripped SN are recognized as core-collapse SN whose massive progenitors have been stripped of progressively increasing amounts of their hydrogen and helium layers (e.g., \citealt{podsiadlowski93,nomoto93,clocchiatti97,matheson01,maurer10}, but see \citealt{dessart12,hachinger12} and below), either
through strong winds \citep{woosley93} or binary interaction (\citealt{nomoto95,podsiadlowski04}) or a combination of both. 

Positive identification of the progenitors of the stripped SN requires the identification of the pre-supernova star and its disappearance after the supernova fades. Recent reviews \citep{smartt09,eldridge13} show that pre-explosion HST images at the sites of SN Ib, SN Ic, and SN Ic-bl have not yet conclusively detected a candidate for the progenitor star (except perhaps for a recent SN Ib; \citealt{cao13}).

Most recently, the connection between long-duration GRBs and
broadlined SN Ic (see \citealt{woosley06_rev,hjorth11,modjaz11_rev} for reviews) has sparked interest in 
the nature and progenitors of SN Ic with and without GRBs. Broad-lined SN Ic are a class of
SN Ic that exhibit very broad and blended lines, indicating significant amount of mass at very large expansion velocities (15,000 $-$ 30,000 \kms)
\citep{galama98,patat01,mazzali06_06aj,modjaz06,sanders11}. The fundamental question remains,
however, whether GRB-SN Ic are a separate class of SN Ic-bl requiring
specific conditions (e.g., low metallicity; \citealt{yoon05,woosley06_z,stanek07,modjaz08_Z}), and/or constitute the extreme end of a continuum of
SN Ic events, dictated by, for example, viewing angle effects in asymmetric explosions.

Despite their importance, there are only a handful of well-studied stripped SN. The best-studied SN Ic has been SN~1994I (e.g.,
\citealt{filippenko95,richmond96}) which exploded in the nearby galaxy
 M~51. Thus, it was well observed over many wavelength regimes and is
commonly referred to as the "proto-typical'' SN Ic. The same applies to SN~1993J (e.g., \citealt{filippenko93,matheson00_93j,matheson00_93jdetail}) in the nearby galaxy M~81, the first SN IIb recognized as such (e.g., \citealt{nomoto93}), and often called the `proto-typical'' SN IIb. To evaluate how typical these SN are for their classes, however, one needs a broad sample for comparison.

The only comprehensive observational spectroscopic study has been conducted by \citet{matheson01},
who presented 84 spectra of 28 stripped SN and unfiltered
light curves for four of their SN Ib. While their spectroscopic
observations are a valuable source for understanding the distinctions
between SN Ib and Ic, their paucity of light curves renders it
impossible to assign a robust phase to most of their spectra.

Understanding the full range of massive stellar explosions requires the study of a large and comprehensive SN sample with homogeneous, and densely time-covered data. It is important to obtain secure spectroscopic identifications that may accurately indicate the outer composition of the pre-explosion star, including very early-time spectra that probe the outermost layers. Such data are presented in this work, which constitutes the largest spectroscopic dataset of stripped SN to date. 

Since 1993, a vigorous program has been conducted at the CfA\footnote{http://www.cfa.harvard.edu/supernova/}
of following up supernova discoveries with photometry and spectra from
the Fred Lawrence Whipple Observatory (FLWO) on Mount Hopkins,
Arizona. Early work emphasized SN Ia for their cosmological utility.
But more recent work, since 2004, has included energetic studies of
stripped-envelope core-collapse supernovae. Photometric results on SN Ia are available in \citet{riess99} (CfA1), \citet{jha06} (CfA2), \citet{hicken09} (CfA3) and \citet{hicken12} (CfA4).  Infrared observations of Type Ia supernovae from
the CfA are available in \citet{wood-vasey08} and in Friedman et al.
(in prep). A first set of 432 CfA spectra of 32 SN Ia was published by \citet{matheson08}, and a recent sample comprising 2603 CfA spectra of 462 SN Ia was published by \citet{blondin12}. All published CfA spectra are publicly available through a
CfA Supernova Archive\footnote{http://www.cfa.harvard.edu/supernova/SNarchive.html}. 

Here we present the aggregate collection of
spectroscopic data of nearby ($z_{\rm{mean}} = \snzmean \pm \snzstdev $)
stripped-envelope SN collected between 1994 and 2009 (with 78\% of all spectra taken during 2004-2009). 
The plots of the spectra for all the SN in this paper are available in the on-line edition, and the data will be made available through the above-mentioned CfA Archive. 
A future companion paper presents the photometric data of these stripped SN (F. B. Bianco et al, in prep). A number of the CfA spectra of a few of these stripped SN, especially historic SN, were published as single objects, as well as part of slightly larger studies (see below for full list), but we include those here, since in most cases, the spectra were re-reduced to be consistent with the full sample. Even for SN that have had spectra published by other groups, the spectra presented here sometimes cover earlier epochs
that were not reported by others. The consistency
of the method of observation and the technique of reduction makes this an important data set for studying the
spectroscopic diversity of stripped SN, as well as providing a comparison sample for SN associated with GRBs, all of which which we undertake in our follow-up paper. We incorporated the data into the Supernova Identification (SNID) code \citep{blondin07} to aid in the (re)classification of SN, as discussed in \S~\ref{id_sec} and \S~\ref{diffid_sec}.

The format is as follows: In \S~\ref{specred_sec} we describe our spectroscopic observational and reduction techniques.  In\S~\ref{spec_sec}  we present
our full data set of optical
spectra.  In \S~\ref{sample_sec} we introduce our stripped SN sample and give background
information about their discoveries and host
galaxies, while discussing in detail the SN classifications in \S~\ref{id_sec} and \S~\ref{diffid_sec}.We discuss individual SN, for which data have already been published, in
\S~\ref{notes_sec}, and conclude with \S~\ref{conclusions_sec}. All \ntotalspec\ spectra presented in this paper will be made publicly available through the
CfA Supernova Archive\footnote{http://www.cfa.harvard.edu/supernova/SNarchive.html}.

\section{Spectroscopic Observations and Data Reduction}\label{specred_sec}

While pre-2004, only the closest and therefore brightest stripped SN were followed, an intense follow-up program started in 2004 (see \citealt{matheson08} and \citealt{blondin12} for detailed discussions of the CfA SN follow-up program criteria).  Our decision to monitor a particular
newly-discovered SN IIb/Ib/Ic was broadly informed by the following
considerations: 1) accessibility (northern SN with declination $\ga$
$-$20$\arcdeg$), 2) SN brightness ($m <$ 18 mag for spectroscopic
observations and $m <$ 20 mag for optical photometry) and 3) SN phase
(SN whose spectra indicated a young age were given higher
priority). Of course 2) and 3) are correlated post maximum, since older SN are
dimmer.

The nearby SN we monitored were discovered by a variety of
professional SN searches and amateurs using modern CCD
technology. Systematic SN searches include the Lick Observatory SN
Search (LOSS, \citealt{filippenko01}), the Texas SN
Search\footnote{http://grad40.as.utexas.edu/~quimby/tss/index.html}
\citep{quimby06_thesis} and the Nearby SN Factory \citep{snfactory02}, amongst others.  The successful LOSS and many amateur SN searches possess a 
relatively small field-of-view (FOV, $8{\farcm}7 \times 8{\farcm}7$
for LOSS) and have a database of galaxies that they monitor
nightly. Thus, they include well-known and inevitably, more luminous
galaxies (e.g.,
\citealt{li01,gallagher05,mannucci05}). SN which were discovered because their host galaxies were specifically targeted by traditional searches are designated as "targeted". The Texas SN Search and the Nearby SN Factory are rolling searches that scan wide areas that include
hundreds or thousands of galaxies without selecting any individual
galaxies as targets. These galaxy-impartial searches are more likely
to include supernovae that erupt in low-luminosity hosts.  It is
important to keep track of the provenance of each supernova to track
selection effects that arise from the search method \citep{modjaz08_Z,young08,sanders12}. Supernovae that
are discovered in galaxy-impartial searches or that are discovered in
background galaxies that are not the target of searches from a list,
we label "untargeted". In our sample, 78\% of the events were discovered in targeted surveys.
Of those targeted SN, 39\% were (co-)discovered by LOSS and 58\% by amateurs, where we have counted the SN twice if they were co-discovered by both LOSS and amateurs independently.

Our data acquisition and reduction techniques are in most part identical to those discussed in \citet{matheson08} and in \citet{blondin12}, and therefore, we only briefly summarize their main aspects here. 
Most of the early-time spectroscopic SN monitoring was performed via queue
observers at the 1.5m Tillinghast telescope at FLWO that produced the
vast majority (\ntotalspecfast\ of \ntotalspec\ spectra) of the
low-resolution spectra presented here. 
In addition, we observed SN
for late-time studies ($\ga$ 3-6 months) with the 6.5 m Clay Telescope
of the Magellan Observatory located at Las Campanas Observatory (LCO),
the 6.5m MMT at FLWO  and the 8.1m
Gemini-North Telescope in Hawaii via queue observations (Program IDs GN-2005B-Q-11 \&
GN-2006B-Q-16, GN-2007A-DD-4, GN-2008A-Q-17, PI: M. Modjaz). The spectrographs utilized were the
FAST \citep{fabricant98} at the FLWO 1.5m telescope, the LDSS-3
(Mulchaey \& Gladders 2005) at LCO, the Blue Channel
\citep{schmidt89} at the MMT and GMOS-North \citep{hook03} at Gemini-North.

Observational details of the spectra and their reductions are given in Table~\ref{specobs_table} in Appendix~\ref{app_sec}. The FAST spectra were obtained in the standard COMBO-setup (slit
width of 3\arcsec, 300 line/mm grating), with an observed wavelength
range of 3500\AA\ $-$ 7400\AA. The wavelength resolution was $\sim$ 7
\AA, in other words, $R = \lambda / \Delta
\lambda \sim$  800 at the central wavelength of 5600\AA, corresponding to a velocity resolution of $\sim$ 400 \kms. During bright time, the FAST 
spectrograph is not mounted on the FLWO 1.5m; hence, some of our
spectroscopic time coverage has gaps of 10 or more days. Moreover, FLWO undergoes a shutdown during the month of August every year, due to Arizona monsoon season, thus, we do not have monitoring data 1 month out of the year. 
For MMT+Blue Channel,
we used 1\arcsec\ or 2\arcsec\ slits and the 300 line/mm grating and for
Gemini+GMOS, the $0{\farcs}$75 slit and 400 line/mm grating. For LDSS3+Magellan, depending
on the availability, we used the VPH-Blue, VPH-Red or VPH-All
grisms, and usually slits of $0{\farcs}$75,
1\arcsec, and $1{\farcs}25$ (matching the seeing conditions). For all
setups, the resolution was 5$-$10 \AA, as determined from the width of
night-sky lines. Spectrophotometric standard stars (e.g.,
\citealt{massey88,hamuy92}) were observed on the same nights for flux
calibration (see Table~\ref{specobs_table}) and are also used to
remove telluric absorption features from the spectra (see, e.g., Wade \& Horne 1988; Matheson
et al. 2000). Since the FAST spectrograph suffered from second-order
light contamination during our observations, we observed standard
stars of different colors. By calibrating the data with blue standards
for the blue wavelength range and with red standards for the red, we
minimized the impact of the second-order light \citep{matheson08}. All optical spectra
were reduced and calibrated employing standard techniques in IRAF and
our own IDL routines for flux calibration (see e.g.,
\citealt{matheson08}). Table~\ref{specobs_table} gives the slit
position angle for each spectrum as well as the absolute difference, $|\Delta\Phi|$, with the
parallactic angle. From 2000 on, after which most of the discussed spectra were taken, the spectra were always taken at or near the parallactic angle \citep{filippenko82}. For the CfA SN Ia samples taken with the same instruments, the relative flux
calibration has been demonstrated to have a precision of $\sim 8\%$ across 4000$-$6500 \AA\ \citep{matheson08,blondin12}.

Here we check the accuracy of the relative flux calibration of our stripped SN spectra, employing a similar procedure as for the CfA SN Ia data in \citet{matheson08} and in \citet{blondin12}. As outlined in Appendix~\ref{flux_sec}, stripped SN have a scatter of 0.24 mag (including all phases), about twice as large as SN Ia from the CfA SN Ia data, using our re-analysis.

%%%%%%%%%%%%%%%%%%%%%%%%%%%%%%%%%%%%%%%%%%%
\begin{figure*}                
%%\hskip-1.cm 
\includegraphics[scale=1.0,angle=0]{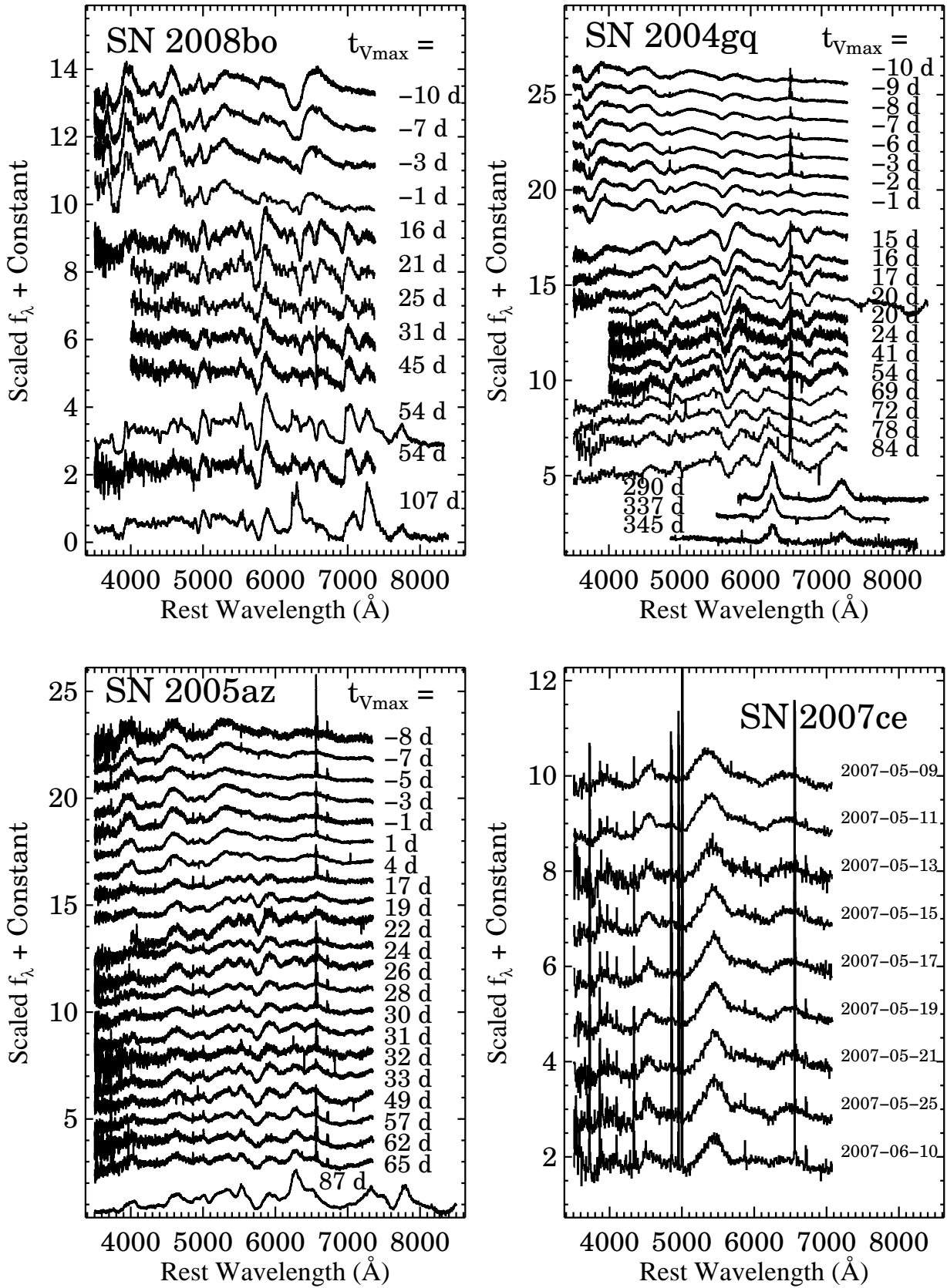}

\singlespace \caption{Example of CfA spectra of one SN per stripped SN type: SN 2008bo (IIb), 2004gq (Ib), 2005az (Ic), 2007ce (Ic-bl).
The flux units are in f$_{\lambda}$ (\ergscmA) that have been
 normalized with additive offsets for clarity. The observation dates
 are indicated with respect to the $V$-band maximum light, $t_{\mathrm{Vmax}}$ (see Table~\ref{snsample_table} for SN with measured date of maximum and corresponding references). For SN without known date of maximum light (here e.g., 2007ce), we list the UT dates of observations instead of phases.
 The spectra, some of which are binned, are in the rest frame of the supernova's host galaxy, as listed in Table~\ref{snsample_table} and described in the text.}
\label{ch4_expmont_fig} 
\end{figure*}                                                                 
%%%%%%%%%%%%%%%%%%%%%%%%%%%%%%%%%%%%%%%%%%%

\section{CfA Stripped SN Spectra}\label{spec_sec}

We present a total of \ntotalspec\ spectra of \nsnspec\ SN in this
section.

Since all \ntotalspecfast\ spectra from the 1.5m Tillinghast were obtained with the same instrument and telescope
set-up, and were reduced in the same manner, our data set constitutes
the largest homogeneous and densely time-sampled spectroscopic dataset of stripped-envelope SN to date. The homogenous aspect of our data is important for our ability to conclusively distinguish between SN Ib and SN Ic, as we are able to consistently test for the presence of the important line of \HeSeven\ (see \S~\ref{id_sec}) at the wavelengths where some of the telluric lines are located, which we remove from all SN spectra in the same manner.
In addition, the MMT and Magellan spectra were also obtained and reduced in the same manner as the data obtained with the 1.5m Tillinghast (e.g., observed at parallactic angle after 1993, telluric absorption removed etc). In Figure~\ref{ch4_expmont_fig}, we only show the CfA spectral series of four example  SN, namely one SN per stripped SN class (SN IIb, Ib, Ic, and Ic-bl). This is done for brevity  and for guidance regarding their format. The rest of the spectra-plots are available in the online version of the paper and will be made available on the CfA SN Archive website, along with the spectral data.

%%%% FIGURES %%%
In all SN spectra figures
%(Figures~\ref{ch4_94imont_fig}$-$\ref{ch4_06ld06lv_mont_fig}) 
and in the
subsequent analysis, the heliocentric SN velocity has been removed in
order to place the spectra in the SN rest frame. The SN velocity was
either determined from narrow emission lines in the galaxy/SN spectrum (from
superimposed \ion{H}{2} regions) when detected, or assumed to be equal
to the heliocentric recession velocity of the host galaxy nucleus
(Table~\ref{snsample_table}), with an uncertainty of up to $\pm$ 200$-$300 \kms\ due to host galaxy's rotation curve . Furthermore, for SN for which the date of maximum light has been measured (either in the literature or in our companion light curve paper by F. B. Bianco et al, in prep, see Table~\ref{snsample_table}), we list the rest-frame phases (i.e,. $\Delta t /(1+z)$) of the spectra, where $\Delta t$ denotes days with respect to maximum light. For most of our SN, maximum light is measured in the $V$-band, except for \nsnrmax\ SN which only have measured $R$-band maxima. For SN without observed date of maximum light, we list on the figures the UT dates of the obtained spectra. 
%\clearpage
%%%%%%%

%%%%%%%%%%%%%%%%
%%%% SN SAMPLE %%%%%

\section{CfA Stripped SN Sample}\label{sample_sec}

We list our SN sample along with pertinent SN and host galaxy information in Table~\ref{snsample_table}.  The SN type listed in column 2 is based on our own assessment by running the SN identification code SNID \citep{blondin07} with its extensive library of template spectra (augmented with more recent spectra than available when published in 2007) on the multi-epoch CfA spectra, which lead in a number of cases to a different SN classification than announced in the IAUC, as we discuss in detail in \S~\ref{id_sec}. With some spectra of SN, it may be hard to tell whether the optical He optical lines are present - those SN we designate as SN Ib/c (for a detailed discussion of the SN classification see \S~\ref{id_sec}). Column 3 lists the discovery reference as the IAUC or CBET number, while column 4 notes the mode in which the SN were discovered ("targeted" vs. "untargeted"). The mode of discovery will be important for environmental studies, specifically for metallicity studies (e.g., see review by \citealt{modjaz11_rev}). Columns 5, 6 and 7 list the host galaxy name and the SN offsets from the galaxy nucleus, which are taken from the
International Astronomical Union Circulars (IAUC) and Central Bureau Electronic Telegrams (CBET)\footnote{http://cfa-www.harvard.edu/iau/cbat.html} and from the NASA/IPAC Extragalactic Database
(NED).  Column 8 lists the heliocentric recession velocity, while column 9 lists the date of maximum light in the V-band, unless noted otherwise, as well as the corresponding reference publication in parenthesis. The CfA values for the dates of maximum values are based on the fits to the CfA photometry using Monte Carlo simulations (F. B. Bianco et al in prep.)

We consulted the Updated Zwicky Catalogue \citep{falco99} and the SDSS catalogue for the 
heliocentric host galaxy redshifts ($cz_{\rm{helio}}$), as well as already published ones for the SN in our sample, and
otherwise used those listed in the NED, giving preference to optical redshifts over \ion{H}{1}
redshifts, if there was a discrepancy. Some of the listed redshifts were determined by ourselves via the superimposed \ion{H}{2}-region emission lines in the SN spectra, if there were no optical redshift determination for the host galaxy (for SN 2005kf, 2007ag, 2007ce) or if the systemic galaxy recession velocities listed in NED were different from those at the SN position by more than $\pm$ 200 \kms\ presumably because of the galaxy's rotation curve (for SN 2004gq, 2005az, 2005bf, 2005la, 2006T, 2006lv).

The last column lists the Julian Date of $V$-band maximum (but in \nsnrmax\ cases, $R$-band maximum, as noted in the table) or upper limits on the date , as well as the appropriate references. From a sample of well-observed $V$ and $R$ light curves of SESN  (e.g., \citealt{drout11} and references therein), it appears that the $R$-band maximum occurs 1$-$3 days later than $V$-maximum.

A number of the CfA spectra of the following SN were published as single objects: SN 1994I \citep{millard99}, 1997ef \citep{iwamoto00}, 2000H \citep{branch02}, 2003jd \citep{valenti08}, 2005bf \citep{tominaga05,modjaz08_doubleoxy}, 2006aj \citep{modjaz06,modjaz08_doubleoxy} and 2008D \citep{modjaz09}. %and 2011dh (H. Marion, in prep).
Furthermore, a sample of CfA late-time spectra of stripped SN were published and analyzed in \citet{modjaz08_doubleoxy} and \citet{milisavljevic10}. While a few SN have been observed by others and spectra published elsewhere (e.g., SN 1998dt, 2002ap, 2004aw, 2005la, 2006jc, 2007gr, 2008ax, 2009jf), we are here contributing new and hitherto unpublished CfA data for those SN. Furthermore, even for a number of SN (SN 1998dt, 20000H, 2004aw, 2005la, 2007bg, 2008ax) that have had spectra published by other authors, our spectra presented here start earlier than any of the published ones . Also, the CfA spectra for some of the SN that have been already published (SN 1994I, 1997ef, 2003jd, 2005bf, 2006aj, 2006jc, 2008D) are amongst the earliest spectra in the literature, which emphasizes the importance of the CfA SN follow-up program and its legacy value. We discuss the publication context of these SN individually in \S~\ref{notes_sec}.

\subsection{SN Type Identification}\label{id_sec}

\begin{figure}[!ht]                
%%\hskip-1.cm 
\includegraphics[scale=0.45,angle=0]{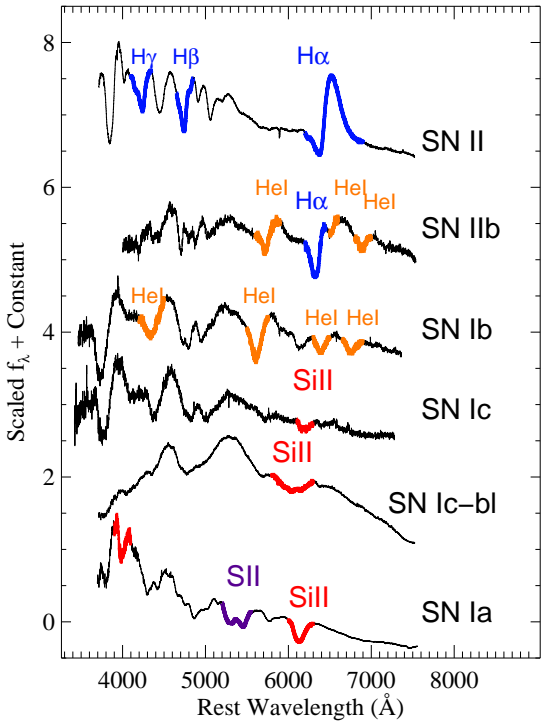} 
\hskip1.cm 

\singlespace \caption{Montage of the main types of stripped SN at around maximum light, along with a SN II and a SN Ia, for comparison, in order to illustrate the SN classification scheme. The main distinguishing spectral features are marked in color and annotated. For clarity, superimposed HII region lines are clipped. The SN plotted are: SN~1999em (SN II, \citealt{leonard01_99em}), SN~1996cb (SN IIb, this work), SN~2004gq (SN Ib, this work), SN~2004fe (SN Ic, this work), SN~2002ap (SN Ic-bl, this work, see also \citealt{gal-yam02_02ap,mazzali02_02ap,foley03}), and SN~1994D (SN~Ia, \citealt{patat96,blondin12}). How much diversity there is within the spectral classification classes and whether there is a continuum between types corresponding to a continuum of envelope masses will be the subject of our follow-up paper that include quantitative comparisons (Y. Liu et al. in prep). Note that some of the identifying spectral features are time-dependent (see text).}
%SN 2006aj/GRB 020618 (SN Ic-bl, \citealt{modjaz06}).
\label{snclassifcation_fig} 
\end{figure}

Given our strength of having obtained multi-epoch spectra for a large number of SN, as well as knowing the date of maximum light such that we can assign phases to the spectra, we now attempt to obtain secure SN type identifications for the SN in our sample. SN are classified based on the presence and absence of specific characteristic elements in their optical spectra at
maximum light \citep{filippenko97_review,matheson01}. For illustration, we plot in Figure~\ref{snclassifcation_fig} some of the main types of SN at around maximum light.  In the following, we adopt the line identifications for
stripped SN of \citet{branch02,elmhamdi06,branch06,dessart12}, as well as those of \citet{baron99,tominaga05,sauer06,valenti08_07gr,tanaka09} for the specific SN that they published and are included in this work. We note that our line identifications are not definitive since we have not conducted spectral synthesis calculations; furthermore, we discuss below some of the features for which contradictory identification exist in the literature.

A number of crucial SN spectral features are heavily time-dependent, making the identification based a single spectrum, upon which the IAUC/CBETS are based, sometimes difficult. For example, for SN IIb, the Balmer lines become weaker and the \He\ lines stronger over time; for SN Ib, the \He\ lines usually emerge over time, due to non-thermal excitation by gamma-rays from the $^{56}$Ni decay chain (for early discussions about the distinction between SN Ib and SN Ic see \citealt{lucy91,wheeler94,clocchiatti96,matheson01}, and more recently, e.g., \citealt{modjaz09,leloudas11,dessart11,hachinger12,dessart12}, and for SN IIb, see \citealt{pastorello08,chornock11,arcavi11,milisavljevic12_IIb}). Given those caveats, some authors question the validity of sub-classifying SN into SN Ib and SN Ic in the first place, and tend to lump the SN subtypes together into the expression "SN Ibc".

%%%%%% TABLE: SN SAMPLE 
%\input{ch4/SNSample_table.tex}
%%%%%%%%%%%%%%%

%%%%%%%%%%%%%%%%%%%%%%%%%%%%%%%%%%%%%%%%%%%
%%%%%%%%%%%%%%%%%%%%%%%%%%%%%%%%%%%%%%%%%%%

We find that high signal-to-noise spectra taken at many epochs with known phases, as we have for this sample, are tremendously useful for arriving at a secure classification. In our follow-up paper that includes quantitative line strength ratios and analysis  (M. Modjaz et al, in prep and Y. Liu et al. in prep), we find that the SN types of SN Ib and SN Ic are useful and separate categories for most of the SN\footnote{There may be even two kinds of SN Ib (Y. Liu et al. in prep) - ones with strong \He\ lines, increasing in strength over time, which include SN 1998dt, 1999dn, 2004ao, 2004dk, 2004gq, 2005hg, 2006ep, 2008D, and ones with weak \He\ absorption lines, which disappear 2 weeks after maximum, such as SN 2009jf (\citealt{valenti11}) and 2002ji (as shown here), but where strong \HeTwomicron\ is detected  (\citealt{valenti11}, H. Marion et al., private communication), securely indicating presence of He.}. The distinction between spectra of SN Ib and SN Ic in the optical is most readily visible when considering not only the strong line of \HeFive, but in the addition at least two other He I lines, namely \HeSix\ and \HeSeven. There are of course additional \He\ lines in the Near Infrared (NIR) that ought to be used, such as \HeOnemicron\ and \HeTwomicron\ (see discussion in \citealt{taubenberger06,modjaz09,hunter09_07gr,valenti11,hachinger12,dessart12}), but these lines are outside the spectral wavelength range of all of our spectra. Usually, authors concentrate on \HeFive, since it is the strongest optical \He\ line, but it has the drawback that it overlaps with Na I D at a similar wavelength (\citealt{sauer06,valenti08_07gr} and references therein). These three main optical He I lines are visible as early as 10 days before maximum light in some SN Ib, and fully developed in all SN Ib by maximum light and at $+$15 and $+$30 days (Y. Liu et al., in prep).  The \He\ lines then disappear when the SN start to transition into the nebular phase; especially,  \HeSix\ and \HeSeven\ become very weak beyond $t_{Vmax}>$ 50$-$70 days. 
%Thus, we recommend to obtain multiple spectra over time and to announce the final SN classification when multiple spectra at maximum light and up to to $+$15 days later have been obtained and classified. 
Nevertheless, any observed continuum between SN types that may correspond to a continuum of envelope masses or of $^{56}$Ni mixing will be discussed in detail in our follow-up papers (M. Modjaz et al. in prep; Y. Liu et al in prep), as well as the approach of quantifying the diversity within the spectral classification classes by constructing average spectra for different SN types and measuring line strength changes over time.

%odd SN Ib: 09jf, 02ji, 04gv, 07uy
Because of the time-dependent nature of the emergence of the \He\ lines, there are three SN Ic (2004eu, 2005kl, 2007cl) out of \nsnIc\ SN Ic in our sample, for which we cannot exclude the potential emergence of He lines during times for which we have no spectra (Table~\ref{snsample_table}). For SN~2007cl, our only spectrum was taken 7 days before V-band maximum, and for SN~2005kl, we have spectra at $t_{Vmax}$=$-4$, $+$70 and $+$155 days. While these spectra do not exhibit \He\ lines and are well-matched by SN Ic spectra via SNID, we cannot exclude the possibility that they may have developed \He\ lines around maximum light and later, since we do not possess spectra at those phases. For 2004eu, we cannot exclude that its spectra were taken before maximum light, since we have no light curve data. Thus, for the majority of the SN Ic in our sample (namely, 16 out of 19 SN Ic), we are confident that they are bona fide SN Ic based on their optical spectra.

Nevertheless, even for "bona fide" SN Ic, recent detailed radiative transfer calculations by \citet{hachinger12} and \citet{dessart12} suggest that the lack of obvious optical lines of \He\ in SN Ic spectra may not necessarily be evidence for helium deficiency in their ejecta. They suggest that if there is not sufficient mixing of \synNi\ , SN may not exhibit optical \He\ lines, but may still have some He in their ejecta (less than 0.06$-$0.14 \sm, depending on mixing and the models). They further suggest that the NIR lines of \He , especially the \HeTwomicron\ line, may be better probes of He content than optical lines, though NIR spectra of SESN are still rare. Further predictions of Dessart et al's model for the case that both SN Ic and SN Ib have the same progenitors, but are only distinguished by the amount of mixing of \synNi will be discussed in Y. Liu et al (in prep).

The question of the SN IIb vs. SN Ib classification is more complex, since in some SN, very early-time spectra long before maximum light are needed to detect strong H$\alpha$ \citep{pastorello08,chornock11,arcavi11,milisavljevic12_IIb}, and these are rare. However, we do have a number of very early pre-maximum spectra of SN Ib, which do not show strong H$\alpha$. For SN 2004gq, 2005hg, 2008D, 2009iz (at $t_{Vmax} \approx -14$ to $-$10 days), and SN 2005bf, 2006ep, 2009jf, 2009er (at around $t_{Vmax} \approx -$ 7 days), a weak absorption line that could be identified with H$\alpha$ is much weaker than \HeFive\ (see Y. Liu et al. in prep), such that we can exclude for those SN (i.e, for at least 8 SN out of \nsnIb\ SN Ib in our sample) a scenario in which much hydrogen might have gone undetected (though for quantitative conclusions radiative transfer calculations are needed).

However, in all of those cases there is a weak absorption at $6100 - 6200$ \AA\ which could be identified with either high-velocity H$\alpha$ \citep{branch02} or with \ion{Si}{2} \citep{tanaka09}.
%Future detailed 

In addition, the burgeoning field of environmental studies have uncovered systematic environmental trends for different subtypes of SN, thus, strongly suggesting that the empirical SN classification scheme appears physically insightful and motivated. Specifically, SN Ib and SN Ic appear to trace differently their respective host galaxy's distributions of H$\alpha$ emission \citep{anderson08,anderson12} and the brightest blue regions \citep{kelly08,kelly12}, with SN Ic having higher correlations with both parameters than SN Ib, and both more than SN II. These observations have been interpreted by these and other authors to indicate higher progenitor masses for more stripped SN, with SN II at the lowest mass end and increasing upwards towards SN Ib and finally SN Ic (see also \citealt{raskin08} and \citealt{leloudas10}), even for the binary scenario (\citealt{smith11_snfrac}). \citet{kelly12} find that the environments of SN IIb and SN Ic-bl seem to be different from their respective cousins, SN Ib, and SN Ic, despite the fact that the SN IIb/SNIb distinction may be debated \citep{milisavljevic12_IIb}. \citet{arcavi10} find that the population of SN harvested via the untargeted survey PTF is different in low-luminosity vs. high-luminosity galaxies such that only SN IIb, Ib and Ic-bl are hosted by low-luminosity galaxies, but no SN Ic. Furthermore, the field of directly measured metallicities is a rapidly developing research area (\citealt{modjaz08_Z,prieto08,anderson10,modjaz11,leloudas11,kelly12,sanders12}, see review in \citealt{modjaz11_rev}), since metallicity is one of the most basic parameters expected to determine the lives and deaths of massive stars \citep{heger03}. While there are indications that the oxygen abundances measured directly at the explosion sites of SN Ib and SN Ic are statistically different, also including those from untargeted surveys (\citealt{modjaz11,kelly12}, but see \citealt{leloudas11,sanders12}), with in turn SN Ic-bl sites having lower oxygen abundances than those of their cousins SN Ic \citep{modjaz11,sanders12}, more data based on larger statistically data sets from the same untargetted survey are needed.

Thus, for the purposes of this paper, we take the SN identifications at face value and use the standard SN spectral identification tool SNID \citep{blondin07} with its extensive library of template spectra (augmented with more recent spectra than available when published in 2007). SNID is based on cross-correlation of the target spectrum with a set of individual "template" spectra of a pre-determined SN type, and we apply it to all our spectra of each SN in our sample to determine the type of the SN, which we then report in Table~\ref{snsample_table}.

\subsection{New or Refined Classifications for Stripped SN}\label{diffid_sec}

%%%%%%%%%%%%%%%%%%%%%%%%%%%%%%%%%%%%%%%%%%%
 \begin{deluxetable*}{lllll}
\tabletypesize{\scriptsize}
\singlespace
\tablecaption{New or Refined Classifications for Stripped SN }
\tablehead{\colhead{SN Name} & 
\colhead{Old SN Type} & 
\colhead{References for Old SN Type} & 
\colhead{New SN Type\tablenotemark{a}  }  & 
\colhead{References for New SN Type} 
} 
\startdata
1990U\tablenotemark{b}   & Ic     & IAUC 5111, \citet{matheson01} & Ib   & this work \\
1995F   &   Ic    & IAUC 6138, SN Catalogues & Ib  & this work \\ 
1997X  &  Ic     &  IAUC 6554   &  Ib & this work \\ 
1997dq  &    Ib  & IAUC 6770, SN catalogues &      Ic    & this work \\
1997dq   &   Ic-bl  & \citet{matheson01,mazzali04}            &     Ic    & this work \\
1997ef   &   Ib?  &  SN catalogues                      &      Ic-bl	& IAUC 6783, 6786, \citet{iwamoto00} \\
2001ai  &     Ic   & IAUC 7605, SN catalogues  &  Ib & this work \\
2002ji   &  Ic  &  IAUC 8025   &  Ib & this work, M. Marion et al. (in prep)  \\
2004ff  & Ic     & IAUC 8428               & IIb    & \citet{leloudas11}, this work \\
2005U & II  &  IAUC 8475         & IIb  & ATEL 431, this work \\
2005az & Ib    & IAUC 8503          & Ic  & \citet{kelly12}, this work \\
2005bf   &   Ic/b  & IAUC 8509, 8520, IAU SN list    &      Ib    & IAUC 8521,8522 \citet{tominaga05,folatelli06} \\
2005eo & Ic     & IAUC 8605, IAU SN list    &      Ia   & this work  \\
2006lc & Ib/c  & \citet{leloudas11} & Ib  & this work \\ 
2006el  &    Ib  &    SN catalogue  & IIb  &   CBET 614,626, this work \\
2006fo & Ic    &  IAUC 8750                    & Ib  &  \citet{leloudas11}, this work \\ 
2007C  & Ib/c  & \citet{leloudas11} & Ib  & this work \\ 
2007D\tablenotemark{c}  &  Ic   & SN Catalogues		    & Ic-bl  &  IAUC 8794 , this work \\
2007bg\tablenotemark{c}  & Ic    & SN Catalogues	 	    & Ic-bl  & IAUC 8834, \citet{young10}, this work \\	
2007ce\tablenotemark{c}  & Ic   & SN Catalogues		    & Ic-bl  &  IAUC 8843, this work \\ 
2007iq & Ic    & CBET 1101 & Ic/Ic-bl		& this work \\
2007kj & Ib/c  & \citet{leloudas11} & Ib  & this work   \\
2007rw & IIb  & CBET 1155  & II & \citet{arora11}, this work \\
2007uy & Ib    & CBET 1191        & Ib-pec & this work, \citet{roy13} \\
2008aq   &   IIb?   &  IAU SN list &   IIb  &    CBET 1271, this work \\
2008bo  &    IIb/Ib & SN catalogues &      IIb    &    CBET 1325, this work  \\
2009er   &  Ib &  CBET 1817        & Ib-pec   & this work \\
2009iz     &  Ib/c    &  CBET 1947     &   Ib        & this work  
\enddata
\tablenotetext{a}{These are the SN types reported in Table~\ref{snsample_table}.}
\tablenotetext{b}{No new data of this SN are presented here, but a re-classification of the spectra from \citet{matheson01} via SNID were obtained with a larger template library.}
\tablenotetext{c}{These SN were announced as SN Ic-bl in the IAUCs, but the SN catalogues list them as SN Ic. See text for details.}
\label{snidchange_table}
\end{deluxetable*}
%%%%%%%%%%%%%%%%%%%%%%%%%%%%%%%%%%%%%%%%%%%

To determine the subtypes of the SN with SNID, we followed classification criteria similar to those of \citep{blondin07}, namely requiring that the SNID rlap\footnote{In SNID, the rlap value is a measure of the strength of the correlation between the best-matching spectrum and the input spectrum.} value be at least 10 and the three best-matching spectra from SNID all be of the same subtype. In cases where we possessed multiple spectra per SN we performed SNID on the highest S/N spectra, but also checked that the ID was consistent for spectra of the same SN at different phases. We find that for a number SN in our sample, our ID either refines or differs from that reported in the IAU and CBET circulars\footnote{We note that the IAU SN list website (http://www.cbat.eps.harvard.edu/lists/Supernovae.html ) might not always include the updated SN type announced in later IAUC/CBETs and usually lists broad-lined SN Ic as SN Ic.} and therefore, from that in some of the catalogues (e.g., the continuously updated Asiago SN catalogue, \citealt{barbon89}, or Sternberg Astronomical Institute Supernova Catalogue\footnote{http://www.sai.msu.su/sn/sncat/}), as well as from the literature. We present a summary of our SN type refinements in Table~\ref{snidchange_table}, and discuss cases below, where our ID differs from those presented in the literature.

For spectra of SN 1990U, as published by \citet{matheson01}, we find better SNID matches to the spectra of SN Ib 2008D, also based on the detection of the \He\ lines in 1999U, and thus, revise its classification to SN Ib, though we do not present any new spectra here.

While SN~1997dq was first typed as a SN Ib (IAUC 6770, and is still listed as a SN Ib in the Sternberg Astronomical Institute Supernova Catalogue), \citet{matheson01} suggested it to be a SN Ic-bl, arguing that its spectra look very similar to those of SN~1997ef, a bona fide SN Ic-bl. This line of argument that was later fleshed out by \citet{mazzali04} , who suggested that both the sparse light curve as well as the spectra of  \citet{matheson01} were both similar to those of SN~1997ef. Thus, many publications since then refer to this SN as a SN Ic-bl (e.g., \citealt{soderberg06_radioobs,guetta07,modjaz08_Z,taubenberger09,maurer10}). However, based on our seven spectra, including those that were taken earlier than by \citet{matheson01}, used in conjunction with SNID, we find better matches with after-maximum spectra of SN Ic than with SN Ic-bl and therefore assign the type "SN Ic" to this SN.

SN~1999ex, for which three optical and NIR spectra were obtained by \citet{hamuy02}, is still being used as an important SN for comparison.  \citet{hamuy02} compared SN~1999ex with SN 1984L, 1987M, and 1994I and concluded that it was similar to a SN Ic but with stronger He lines, or a SN Ib with "moderate strength" \He\ lines, leading them to classify it as "intermediate" SN Ib/c. However, subsequent papers, including those that perform spectral synthesis with SYNOW \citep{branch02,elmhamdi06,branch06}, conclude that the presence of \He\ is solid and identify the line at 6200\AA\ as H$\alpha$, making SN~1999ex a SN Ib (or even an SN IIb). While we do not present any new spectra  of SN~1999ex, here we run SNID on the three spectra of \citet{hamuy02}, which were taken at $\Delta t_{Vmax}$= $-$5, 0, and 9 days. The spectra are matched well by spectra of other SN Ib (e.g., SN 2005bf, 2007Y) as well as a few SN IIb (e.g. SN 2000H, 2008ax, though SN IIb show a double-dip profile at H$\alpha$, while SN 1999ex only shows a single dip). Thus, for the remainder of this paper, we regard SN 1999ex as a SN Ib and do not call it an intermediate type SN Ib/c.

For SN 2002ji, we note that our re-classification from a SN Ic to a SN Ib is supported by the detection of \HeTwomicron\ in its NIR spectra (H. Marion et al. in prep). 

For 2004dk, we note that based on an early spectrum, \citet{harutyunyan08} list it as a SN Ic in their paper (probably since the He lines had not fully developed two weeks before maximum light), while our spectra spanning $t_{Vmax}=$+17$-$50 days, as well as those mentioned in IAUC 8404 announcing the SN Ib classification, show the four optical \He\~lines.

For SN~2004gt, we do not see all optical lines of \He\ in our six photospheric spectra (spanning the phases $t_{Vmax}=16-48$ days, where usually all He lines are apparent in SN Ib) and therefore refine the classification from "SN Ib/c" to "SN Ic". 

In addition, our multi-epoch spectra support the reclassification of two SN as proposed by \citet{leloudas11}:  SN 2006fo from a SN Ic to a SN Ib, as well as the reclassification of SN 2004ff from SN Ic to SN IIb, based on our pre-maximum spectrum which shows strong H$\alpha$ at $V_{Rmax}=-4$ days. However, for three SN, our IDs differ from those suggested in \citet{leloudas11}, namely their reclassification of SN 2006lc, 2007C and 2007kj from SN Ib to SN Ib/c. Their re-classifications are mostly based on the fact that Leloudas et al. found good matches with spectra of SN 1999ex (G. Leloudas, priv. communication), which they call a SN Ib/c following \citet{hamuy02}. For all three of these SN, our own multi-epoch spectra clearly show the multiple optical lines of \He\ (\HeFour, \HeFive, \HeSix\ and \HeSeven ) and the spectra are fit by SNID with a number of SN Ib (SN 1999dn, 1999ex, 2005hg). 

Indeed, when obtaining the spectra of SN 2007C and 2007kj from \citet{leloudas11} and running SNID with our updated set of templates on them, we find in both cases very good matches with clear SN Ib (SN 2005bf, 2008D) besides SN 1999ex, which we find to be a SN Ib (see above). Thus, we classify these SN as SN Ib, in line with the classification based on our own spectra.

 Furthermore, our spectra support the reclassification of SN 2005az in \citet{kelly12} from a SN Ib to a SN Ic based on our 24 spectra (that range in phase from $t_{Vmax}= -8$ to $+87$ days), since not all optical \He\ lines are visible in the spectra that usually are visible in normal SN Ib and since SNID finds better matches with SN Ic (e.g., SN Ic 2004aw, 2007gr) than with SN Ib. The strongest line at $\sim$5800 \AA could be either due to He I $\lambda$5876 for a bona fide SN Ib (if all the other optical He lines were also visible), or due to \ion{Na}{1} D, or due to a blend \citep{baron99,sauer06,valenti08_07gr}.

For SN~2005eo, SNID finds many very good matches with normal SN Ia at 2-3 weeks after maximum, and a few, but worse, matches with young SN Ic (around maximum). Furthermore, the maximum observed in the $R$-band light curve of SN~2005eo in \citet{drout11} is consistent with the second $R$-band maximum (or hump) observed in the light curves of SN Ia (as shown in F. Bianco et al. in prep), making the phases derived from the spectra based on SNID SN Ia template fitting be fully consistent with the spectral phases read off directly from the light curve. Thus, we reclassify this SN from a SN Ic to a SN Ia, remove it from our stripped SN sample, and are reminded that spectra of young SN Ic can look similar to those of older SN Ia (e.g., \citealt{blondin07}).

 We note that while SN~2007D, 2007bg and 2007ce were announced as SN Ic-bl $-$ as confirmed by our spectra and by \citet{young10} for SN~2007bg, they are simply listed as SN Ic on the IAU SN website and in the various catalogues. 
 
While SN 2007iq was classified as a SN Ic based on a  spectrum taken on 2007 Oct. 15 (CBET 1101), our CfA spectrum was taken 1 month earlier (on 2007 Sept 14). Running SNID on our spectrum yields good matches with spectra of both old SN Ic and old SN Ic-bl. Thus, we conclude that a unique identification of this SN based on spectra presumably taken well after maximum light is not secure and reclassify this SN from a SN Ic to a SN Ic/Ic-bl. 
   
 For the four spectra of SN2007rw, SNID finds better matches with SN II spectra (both SNIIP and SNIIL) than with SN~1993J, a SN IIb, and we see strong H$\alpha$ in its nebular phase spectra. Thus, we revise its SN type from IIb to II, in agreement with \citealt{arora11}, and remove it from our stripped SN sample.
 
\begin{figure}                
%%\hskip-1.cm 
\includegraphics[scale=0.4,angle=0]{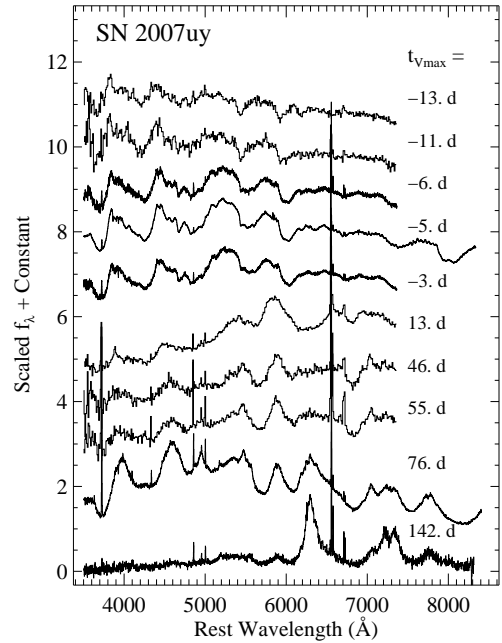} 
\singlespace \caption{Time evolution of the CfA Spectra of SN Ib-pec 2007uy, as in Fig.~\ref{ch4_expmont_fig}. }
\label{sn07uymont_fig} 
\end{figure}

We present our spectral time series of SN~2007uy in Figure~\ref{sn07uymont_fig}. While SN~2007uy was initially classified as a SN Ib  by us based on our earliest spectrum at $t_{Vmax}=-13$ days \citep{blondin08_07uyid}, its subsequent evolution is unlike that of any observed SN Ib, except one. The spectral feature that could be identified as \HeFive\  is not as strong as usually seen in SN Ib, and the absorption line at $\sim$6100 \AA\ is very strong, much stronger than observed in both SN Ib and SN Ic. Indeed, SNID finds many matches with SN Ia (where the line at $\sim$6100 \AA\ is consistent with the strength of \SiSix\ in SN Ia) and only few matches with SN Ib at later epochs (around and within two weeks after maximum, including SN~2009er, see below). Nevertheless, there are reasons to believe that SN~2007uy is a SN Ib, albeit a very peculiar one, as we discuss below. In any case, SN~2007uy is a H-stripped core-collapse object, since its nebular spectra unambiguously show the presence of intermediate-mass elements (Oxygen, Calcium), and no H, which is characteristic of stripped core-collapse SN, with line profiles consistent with other SN Ib (see also \citealt{milisavljevic10,roy13}). Its nebular spectra do not show the iron blends that are characteristic of SN Ia nebular spectra.

%%%%%%%%%%%%%%%%%%%%%%%%%%%%%%%%%%%%%%%%%%%
%% made with idl/mspec/mkmont_dates.pro and with phaselist made by mkjd.pro
%%%%%%%%%%%%%%%%%%%%%%%%%%%%%%%%%%%%%%%%%%%

%%%%%%%%%%%%%%%%%%%%%%%%%%%%%%%%%%%%%%%%%%%
\begin{figure}                
%%\hskip-1.cm 
\includegraphics[scale=0.4,angle=0]{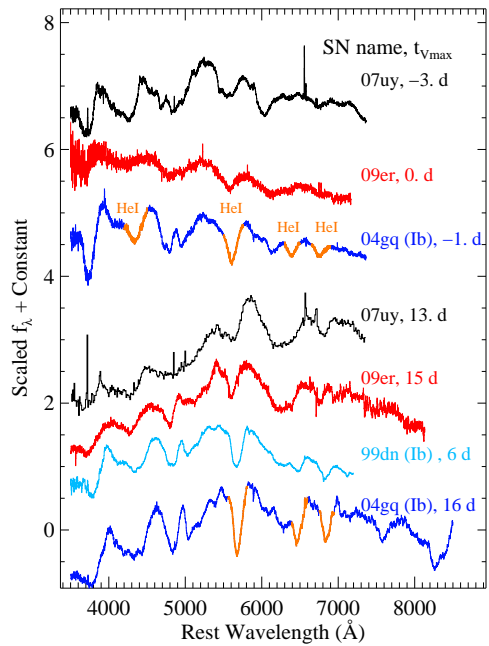} 
\singlespace \caption{Comparison of spectra of SN~2007uy with those of other SN Ib at maximum light and at $t_{Vmax}$= 2 weeks. We show the spectra of SN~2004gq (in blue), a classical SN Ib, at 2 epochs closest to those of SN~2007uy, and color code the He lines with orange to guide the eye. We also show spectra of SN~2009er (from this work) which has a very similar spectrum to SN~2007uy, at 2 weeks after maximum. Data shown for SN~1999dn include those from \citet{benetti11}. }
\label{sn07uycomp_fig} 
\end{figure}                                                                 
%% made with idl/mspec/mkmont_dates.pro and with phaselist made by mkjd.pro
%%%%%%%%%%%%%%%%%%%%%%%%%%%%%%%%%%%%%%%%%%%

\begin{figure}                
%%\hskip-1.cm 
\includegraphics[scale=0.4,angle=0]{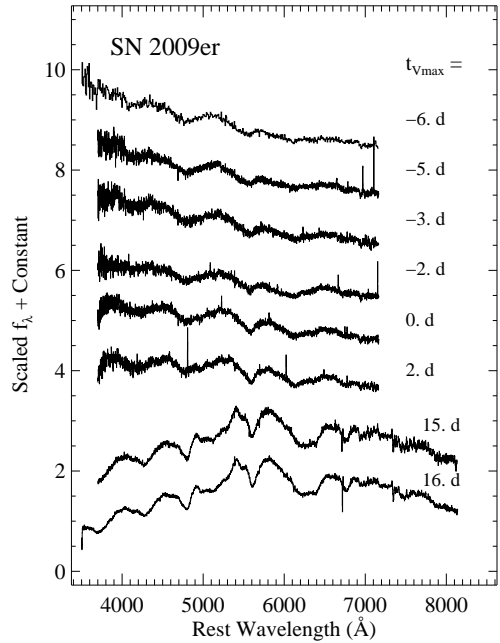} 
\singlespace \caption{Time evolution of the CfA Spectra of SN Ib-pec 2009er, as in Fig.~\ref{ch4_expmont_fig}.}
\label{sn09ermont_fig} 
\end{figure}

In Fig~\ref{sn07uycomp_fig}, we plot spectra of SN~2007uy at maximum and at two weeks after, i.e., phases were the Helium lines are well-developed in normal SN Ib, along with spectra of SN Ib 2004gq, a normal SN Ib, and SN 2009er, at the same phases. The normal SN Ib 2004gq clearly shows the set of three Helium lines, while SN~2007uy shows weak absorption at wavelengths consistent with those three Helium lines at somewhat higher velocities than observed in other SN Ib (by $\sim$ 1000-2000 \kms). 
For the spectrum at $t_{V_{max}}$=13 days, if the lines at $\sim$5500 \AA\ and 6100 \AA\ are due to \HeFive\ and \HeSix\  (respectively), they would have to be blended with other lines to explain their observed large widths (15,000 \kms and 30,000 \kms, respectively) in SN~2007uy. Indeed, the strong, unidentified line that was at 6000-6100\AA\ between $t_{V_{max}}$=$-$6  and $-$3 days may have merged with \HeSix\ on $t_{V_{max}}=$13 days to give rise to the very wide absorption line - unfortunately we have no data during that observing gap to witness and test this development. We have found one other SN Ib with similar spectra at 2 weeks after maximum, namely SN~2009er, which appears to be a case of an in-between object between normal SN Ib (such as SN2004gq and SN1999dn, the later having a notable, but less deep absorption at 6200-6300 \AA) and the extreme case of SN~2007uy. SN~2009er shows on $t_{V_{max}}$=15 days  the same broad line at $\sim$ 6100 \AA like SN~2007uy (which is seen clearly in Figures~\ref{sn09ermont_fig} and \ref{sn07uycomp_fig}), but in the blue part of the spectrum, SN~2009er has a less smeared out spectrum, i.e. the lines are less blended than in SN~2007uy; for example, the blue edge of the line that may be identified with \HeFive\ is more well-defined in SN~2009er than it is in 2007uy, leading to many more matches with normal SN Ib via SNID for SN~2009er than for SN~2007uy.  While it outside the scope of this paper to conclusively identify the deep feature at 6100 \AA, we note that \citet{benetti11} have an extensive discussion of the similar, but much weaker and narrow feature in SN~1999dn, which has been attributed by different authors using different spectral synthesis codes to either H$\alpha$ \citep{branch02,james10,benetti11}, CII \citep{deng00} or a blend of \SiSix\ and FeII lines \citep{ketchum08}. Lastly, in SN2007uy the noted strong absorption line at $\sim$ 6100\AA is not similar to the strong H$\alpha$ line seen in SN IIb, as SNID does not get any good matches with SNIIb for SN 2007uy and thus SN2007uy cannot be called a SN IIb. When SN~2007uy was re-observed on $t_{V_{max}}=$ 46 days and after, that line is no longer detected in the spectra.

Thus, we denote SN~2007uy as a "SN Ib-pec", i.e, a peculiar SN Ib. Recently, \citet{roy13} presented a spectral series of this SN (including two of our CfA spectra\footnote{We note that the CfA spectrum taken on April 1 2008 of SN~2007uy as published in \citet{milisavljevic10} (labeled "92 days [after discovery]) and used in\citet{roy13} (labeled "+96 days [after explosion]") is accidentally that of SN~2008D, observed at the same time in the same galaxy, and not of SN~2007uy. Here we present the correct CfA spectrum of SN 2007uy (labeled as "t$_{V_{max}}$ = 76 days") in Fig~\ref{sn07uymont_fig}. }).  In large their observations are consistent with ours (except when they refer to the fleeting feature which is fleeting because it a feature in the spectrum of SN~2008D), and they attribute its peculiar spectral behavior to a very aspherical explosion in this SESN.

%%%%%%%%%%%%%%%%%%%%%%%%%%%%%%%%%%%%%%%%%%%
%% made with idl/mspec/mkmont_dates.pro and with phaselist made by mkjd.pro
%%%%%%%%%%%%%%%%%%%%%%%%%%%%%%%%%%%%%%%%%%%

The pre-maximum spectra of SN~2009er are characterized by very broad lines (see Fig~\ref{sn09ermont_fig}), similar to SN Ib 2008D which showed SN Ic-bl like spectra after explosion, i.e., 15 to 10 days before maximum, but developed narrow-line spectra with helium by maximum light (\citealt{mazzali08,modjaz09,malesani09}). Our two spectra at $\Delta t_{Vmax} =$ $+$15 and $+$16 days are well-matched by those of a number of normal SN Ib (1990I, 1998dt, 2004gq) at similar phases, but SN2009er exhibits weaker \He\ line strengths and a stronger absorption feature at 6200$-$6400 \AA\ than other SN Ib, and similar to SN~2007uy (see above and Fig~\ref{sn07uycomp_fig}).  Thus, we call SN~2009er a SN Ib-pec, and caution that its early-time spectra and evolution are unusual, as well as its light curve shape and luminosity (H. Marion, private communication, F. B. Bianco et al. in prep).

%%%%%%%%%%%%%%%%%%%%%%%%%%%%%%%%%%%%%%%%%%%
\begin{figure}[!ht]                
\hskip-1.cm 
\includegraphics[scale=0.4,angle=90]{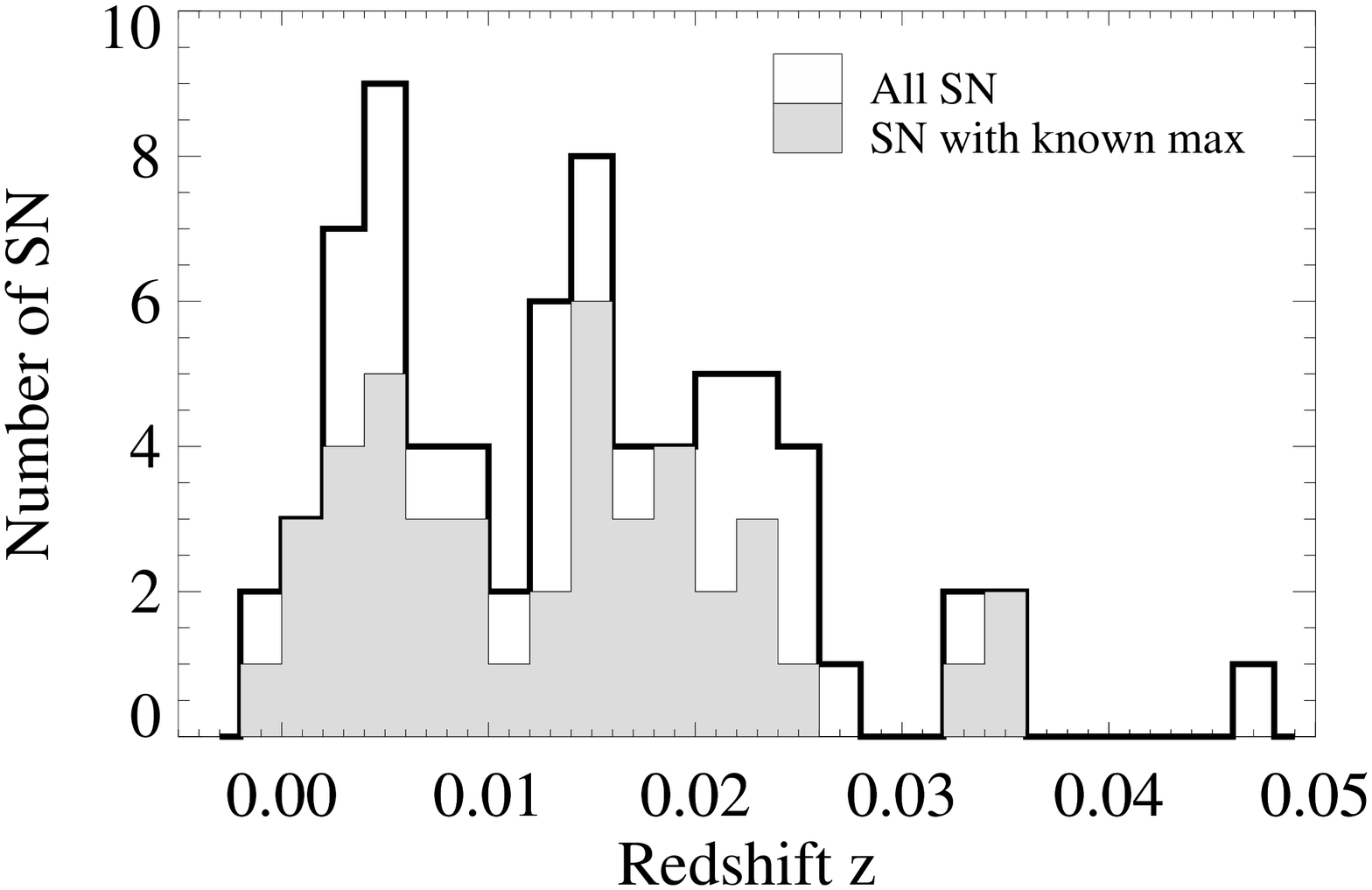} 
\singlespace \caption{Distribution of redshifts of the \nsntot\ SN in the CfA stripped SN sample, including published ones, with a mean redshift value of \snzmean\ $\pm$ \snzstdev . The redshift-distribution of SN with phase information is overplotted in grey. The SN at negative recession velocities are the nearby SN IIb 1993J (at $-$140 \kms) and Ic 2004gk (at $-122$ \kms), while the highest redshift ones consist of three broad-lined SN Ic (\snaj\ at $z$=0.03353, and SN Ic-bl 2007bg at $z$=0.0346 and 2007ce at $z$=0.04633) and one SN Ib (SN 2009er at $z=$0.0350). }
\label{zhisto_fig} 
\end{figure}   
%% MADE WITH /Users/maryammodjaz/Dropbox/AllSNIbc_paper/ch4/cfasnstats.pro 
%%%%%%%%%%%%%%%%%%%%%%%%%%%%%%%%%%%%%%%%%%%

In conclusion, our sample consists of \nsnIIb\ SN IIb, \nsnIb\ SN Ib (where we have included SN Ib-pec 2007uy and 2009er), \nsnIc\ SN Ic, \nsnIcbl\ SN Ic-bl, and  \nsnnotsure\ for SN for which either we could not determine an unique type based on our spectra (SN 1995bb, 2007iq) or which belong to an interacting SN type without much hydrogen (SN 2005la, 2006jc).

%\clearpage
\subsection{SN Sample Properties: Redshifts and Phases}

%%%%%%%%%%%%%%%%%%%%%%%%%%%%%%%%%%%%%%%%%%%
\begin{figure}[!ht]                
\hskip-1.cm 
\includegraphics[scale=0.4,angle=90]{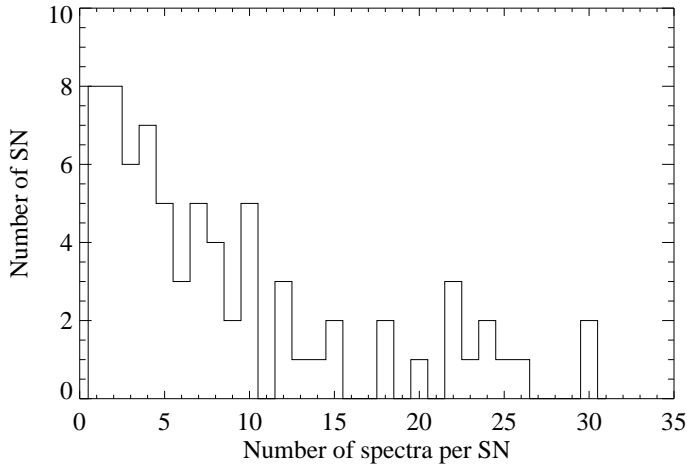} 
\singlespace \caption{Distribution of number of spectra per SN for all \nsntot\ SN in the CfA stripped SN sample. Note that while the maximum number of spectra per SN lies at a value of 2 spectra, there is a large tail of SN with many spectra, such that the average number of spectra per SN for all SN in our sample is \avespecforallsn\  and for SN with observed maximum  \avespecforsnvmax\ (see text for more details).  }
\label{nspechisto_fig} 
\end{figure}   
%% MADE WITH /Users/maryammodjaz/Dropbox/AllSNIbc_paper/ch4/cfasnstats.pro 

%%%%%%%%%%%%%%%%%%%%%%%%%%%%%%%%%%%%%%%%%%%

Figure~\ref{zhisto_fig} presents the redshift distribution of the CfA SESN sample, including published ones, which has a mean redshift of \snzmean\ $\pm$ \snzstdev. The SN at negative recession velocities are the nearby SN IIb 1993J (at $-$140 \kms) and Ic 2004gk (at $-122$ \kms), while the ones at the highest redshifts consist of one SN Ib (SN 2009er at $z=$0.0350) and three broad-lined SN Ic, namely \snaj\ at $z$=0.03353, SN Ic-bl 2007bg at $z$=0.0346 and 2007ce at $z$=0.04633.

%%%%%%%%%%%%%%%%%%%%%%%%%%%%%%%%%%%%%%%%%%%
\begin{figure}[!ht]                
%%\hskip-1.cm 
\includegraphics[scale=0.5,angle=0]{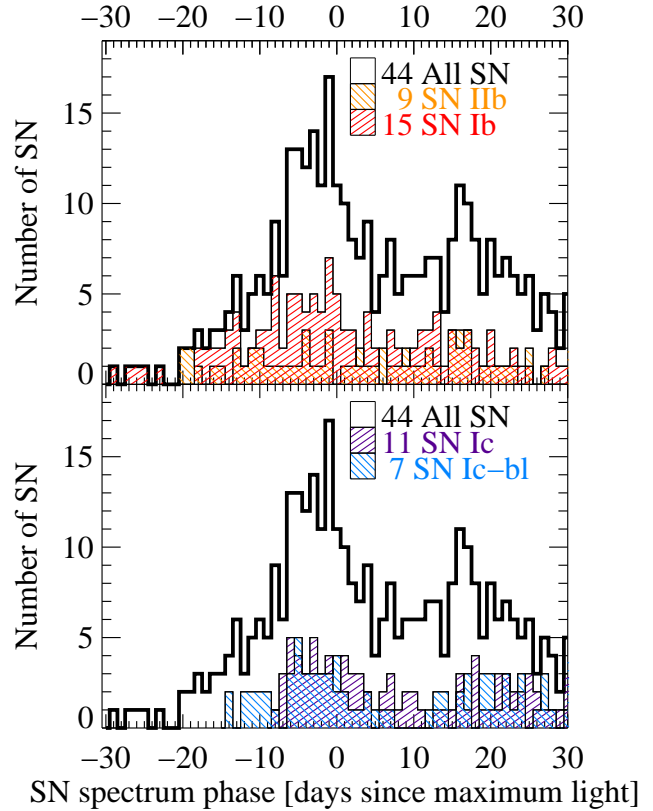} 
\singlespace \caption{Distribution of the spectral phases of the \nsnvmax\ SN in the CfA Stripped SN sample with measured V-max dates, including published ones. The spectral phases earlier than 30 days after maximum are shown, as well as the distribution for different stripped SN sub-types. In this plot, we include SN Ib-pec 2007uy and 2009er amongst SN Ib while SN Ib-n 2006jc and SN Ib-n/IIb-n 2005la are not shown amongst the subtypes, but included in all SN. The number of SN corresponding to a given type is indicated in the legend. Note the large number of pre-maximum spectra for all types, especially for SN IIb and SN Ic-bl.  }
\label{phasehisto_fig} 
\end{figure}
\begin{figure}[!ht]                
%%\hskip-1.cm 
\includegraphics[scale=0.42,angle=90]{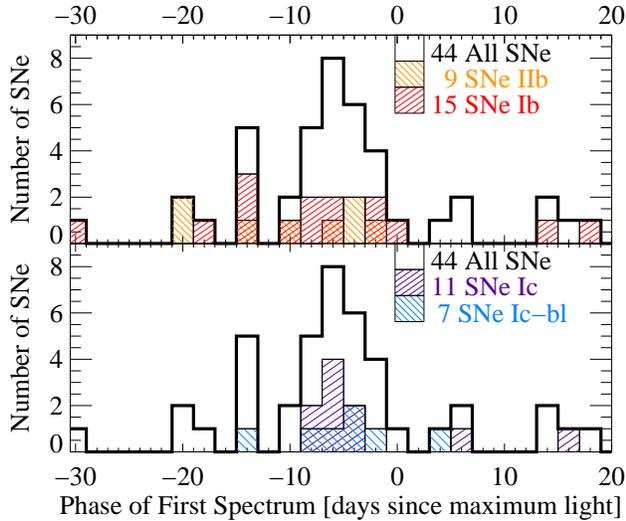} 
\hskip1.cm 

\singlespace \caption{Similar to Figure~\ref{phasehisto_fig}, but now for distribution of phases of the first spectrum. Note the large number of pre-maximum spectra for all types, especially for SN IIb and SN Ic-bl.}
\label{phasehistofirst_fig} 
\end{figure}

   
%%%%%%%%%%%%%%%%%%%%%%%%%%%%%%%%%%%%%%%%%%%
%(regardless of whether maximum was observed or not)
With the full set of \ntotalspec\ spectra for \nsntot\ SN in our sample (see Fig.~\ref{nspechisto_fig} for the distribution), the average number of spectra per SN is \avespecforallsn\ . For the \nsnvmax\ SN with observed light curve maximum, we possess a total of \nspecsnvmax\ spectra, which corresponds to an average of \avespecforsnvmax\ spectra per SN with measured date of maximum light.  Figure~\ref{phasehisto_fig} shows the distribution of the phases of the spectra of  \nsnvmax\ SN in the CfA stripped SN sample, including those of already published ones. The spectra counted here are at a phase earlier than 30 days after maximum, and we plot them also as a function of SN type. Note the large number of pre-maximum spectra for all types, especially for SN Ib and Ic, while SN IIb and SN Ic-bl have the earliest spectra, taken as early as 20 days before $V$-band maximum. We also obtained a number of spectra that are at later spectral phases: for spectra with $30< t_{Vmax}< 60$ days, there are $\sim$ 4 SN spectra for each day (combining all types), with a long tail of SN spectra before and during the nebular phase (not shown). Furthermore, in Figure~\ref{phasehistofirst_fig}, we show the phase of the first spectrum that we took of the SN. Note that the rarer subtypes in our sample (SN IIb and SN Ic-bl) have had their first spectrum taken at earlier phases than those of their more common SN cousins (SN Ib and SN Ic, respectively). We are not using the date of explosion as our reference point for assigning ages to the spectra, since the explosion dates are unconstrained for most of the SN in our sample, except for a few special SN (e.g. GRB-SN, SN 2008D and nearby SN).

%%%%%%%%%%%%%%%%%%%%%%%%%%%%%%%%%%%%%%%%%%%

\section{Notes on Individual SN with Published Early-time Spectra}\label{notes_sec}

Here we list the specific SN for which early-time spectra have been published in the literature (either based on our work or those of others) and relate those published data to our data set presented here. We also note any published late-time spectra, but do not discuss further the nebular spectra SN samples by \citet{maeda08,modjaz08_doubleoxy,taubenberger09,milisavljevic10}.

\vskip0.5cm SN~1993J: This SN IIb in M~81 was a well-observed SN  (e.g., \citealt{filippenko93,trammel93,wheeler93,vandyk94,richmond94,barbon95,matheson00_93j,matheson00_93jdetail,maund04}), and here we present nine very late-time ($t_{Vmax}=$ 590 $-$ 2958 days) CfA spectra on this SN, some of which were included in \citet{milisavljevic12}.

\vskip0.5cm SN~1994I: A number of the spectra of this SN Ic presented here have been published and analyzed by \citet{millard99}, but they were re-reduced to be consistent with the rest of this spectroscopic sample. Since SN~1994I was a bright SN in the nearby M~51, it was well observed by a number of other groups over many wavelength regimes (e.g., \citealt{filippenko95,richmond96,immler98}), leading to its designation as the "proto-typical" or "standard" SN Ic.

\vskip0.5cm SN~1996cb:  This SN IIb was well-observed by \citet{qui99} who presented BVR light curves and 19 spectra that span $-$18 days $< t_{Vmax} <$  154 days. M01 present spectra at later times. Here we present a total of 22 spectra, which start two days earlier than those of \citet{qui99}, and are amongst the  earliest spectra of SN IIb hitherto taken (i.e., 20 days before maximum light), as well as fully nebular spectra taken one year after the explosion.

\vskip0.5cm SN~1997dq:  \citet{matheson01} presented five spectra of this SN Ic.
 Here we present seven spectra of SN~1997dq, which span 7 weeks, amongst which our first spectrum was taken one day earlier than the earliest spectrum presented in \citet{matheson01}.

  \vskip0.5cm SN~1997ef: Some of our early spectra presented here have been published and analyzed by \citet{iwamoto00}, but  the spectra presented herein were re-reduced to be consistent with the rest of this spectroscopic sample. These data                                                                                                                                                                       constitute the most complete spectroscopic data set for this SN that was amongst the first to be typed as a broad-lined SN Ic. Nebular spectra of this SN have been presented in \citet{matheson01} and in \citet{mazzali04}.

\vskip0.5cm SN~1997X: While \citet{munari98} presented two early-time spectra of this SN Ib, here we present
six early-time spectra which start three days before those in Munari et al.

\vskip0.5cm SN~1998dt: While \citet{matheson01} published two spectra of this SN Ib taken with Lick 3m+KAST, which were at  $t_{Rmax}$=$+$8 and $+$33 days, here we present seven new CfA spectra that are at earlier phases between $t_{Rmax}= 0 $ and $+$17 days.

\vskip0.5cm SN~1998fa: Four spectra of this SN IIb were presented in \citet{matheson01} (starting 01-10-99 till 02-12-99), as well as an R-band light curve. Here we present three spectra, which start two weeks before those presented in Matheson et al.

\vskip0.5cm SN~2000H: While \citet{branch02} published spectra based on observations collected at the
European Southern Observatory and at the Asiago Observatory that started after maximum light, here we present seven new CfA spectra of this SN IIb that range in earlier phases from $t_{Vmax}=  -2$ to +17 days. Note that the date of maximum light is uncertain: while \citet{branch02} favor 2000 February 11 as the date of $V$-band max from unpublished ESO photometry, the Asiago Catalogue lists 2000 February 02 as the date of max. Here we adopt the later date consistent with Branch et al., since otherwise the spectral evolution of SN 2000H would be very different from that of the rest of SN IIb, as shown in our companion paper (Y. Liu et al., in prep). 

\vskip0.5cm SN~2001gd: This SN IIb emitted strong radio emission with modulations (e.g., \citealt{stockdale03,ryder04}, see \citealt{stockdale07} and references therein), as well as strong Xray-emission \citep{immler05}. However, since the date of maximum is not measured, the authors use spectral similarity with SN~1993J to estimate its explosion date, which is highly uncertain. Here we present two spectra of this SN which show its transformation into the nebular phase.

\vskip0.5cm SN~2002ap: This SN Ic-bl in M~74 was well observed (e.g., \citealt{gal-yam02_02ap,leonard02,mazzali02_02ap,berger02,foley03}), and here we present 15 new unpublished CfA spectra for this SN.

\vskip0.5cm SN~2003jd: A number of the 24 CfA FLWO spectra presented here have been published and analyzed by \citet{valenti08}, but are shown here for completeness. We note that \citet{mazzali05_03jd} published their two late-time spectra.

\vskip0.5cm SN~2004aw:  While \citet{taubenberger06} published their own spectra of this SN Ic, here we present 15 new CfA spectra, amongst which the earliest one at $t_{Vmax}=-$6 days was taken three days earlier than that in Taubenberger et al.

\vskip0.5cm SN~2004dk: While \citet{harutyunyan08} presented their spectrum taken two weeks before $V$-band maximum light of this SN Ib, here we present three CfA spectra that span $t_{Vmax}=$+17$-$50 days. 

\vskip0.5cm SN~2004dn: While \citet{harutyunyan08} presented their spectrum taken one week before $V$-band maximum light of this SN Ic/Ic-bl, here we present four CfA spectra were taken one month later and that span $t_{Vmax}=$30$-$55 days.

\vskip0.5cm SN~2004fe: While \citet{harutyunyan08} presented their spectrum taken one week before $V$-band maximum light of this SN Ic, here we present 10 CfA spectra were taken at the same time and later, spanning $t_{Vmax}=-6$ to $+242$ days.

\vskip0.5cm SN~2005la: While \citet{pastorello08} published their own four spectra of this peculiar and transitional SN Ib-n/IIn, starting 17 days after R-band max, here we present five CfA spectra that were taken at earlier phases, starting at 14 days after R-band maximum.

\vskip0.5cm SN~2005bf: A number of the CfA spectra presented here have been published and analyzed by \citet{tominaga05}, but  are presented herein with details and for completeness. SN 2005bf had peculiar late-time spectra \citep{maeda07,modjaz08_doubleoxy} and a pronounced and unique double-peaked light curve \citep{tominaga05,folatelli06}. Here we are using the second, main, peak as the peak for V-band maximum, since the time evolution of its spectral features are more in line with those of other SN Ib than otherwise (except the increasing \He\ velocities over time; \citealt{tominaga05}, Y. Liu et al. in prep).

\vskip0.5cm SN~2006aj: SN~2006aj/GRB~060218 was a well-observed SN Ic-bl/GRB \citep{campana06,modjaz06, pian06,sollerman06,mirabal06,cobb06,kocevski07,mazzali07_06aj,taubenberger09}. Our CfA spectra were presented in \citet{modjaz06} and in \citet{modjaz08_doubleoxy}, and are included here for completeness.

\vskip0.5cm SN~2006lc: While \citet{oestman11} presented their spectrum taken on 2006-12-13 of this SN Ib, here we present four CfA spectra were taken two months prior and that span 2006-10-26 to 2006-10-31.

\vskip0.5cm SN~2006fo: While \citet{oestman11} presented their three spectra of this SN Ib, here we present three CfA spectra 
that range from $t_{Vmax}= -3$ days to +75 days.

\vskip0.5cm SN~2006jc:  SN~2006jc was a well-observed SN since it was peculiar SN Ib with He lines in emission, indicating interaction between the SN ejecta and a dense helium shell (e.g., \citealt{foley07,pastorello07,immler07}) and early onset of dust production (e.g., \citet{smith08_dust}. Here we present 19 unpublished CfA spectra that are between $t_{Rmax}= +6$ and +66 days. 

\vskip0.5cm SN~2007bg: While \citet{young10} presented their four spectra of this SN Ic-bl starting at $t_{Vmax}= +$5 days, here we present one spectrum that was taken at  $t_{Vmax}= +$4 days.

\vskip0.5cm SN~2007gr: SN~2007gr was a well-observed SN Ic (e.g., \citealt{valenti08,hunter09,tanaka08_07gr}) in the nearby NGC~1058, with which progenitor detection was attempted \citep{crockett08_07gr}. While \citet{paragi10} claimed it to be an engine-driven explosion without an observed GRB, \citet{soderberg10_07gr} argue based on their radio light curves and X-ray data that it may well be an ordinary SN Ic explosion. Here we present 17 new CfA spectra that range from $t_{Vmax}= +7$ days to +69 days.

\vskip0.5cm SN~2007ru: \citet{sahu09} presented their own nine spectra of this SN Ic-bl starting at $t_{Vmax}= $-2 days, and here we present three spectra that start at $t_{Vmax}= -2$ day.

\vskip0.5cm SN~2007rz: Our CfA late-time spectrum of SN Ic 2007rz was presented in \citet{milisavljevic10}, and here we also present one early-time spectrum. The late-time spectra is contaminated by host galaxy light, especially on the blue end.

\vskip0.5cm SN~2007uy: While one late-time CfA MMT spectrum of this SN was presented and analyzed in \citet{milisavljevic10}\footnote{We note that the CfA spectrum of SN~2007uy shown in \citet{milisavljevic08} (labeled "92 days [after discovery]) and used in \citet{roy13} (labeled "+96 days [after explosion]") is accidentally that of SN~2008D and not of SN~2007uy.}, we present here the additional nine early-time spectra of this SN, which range from $t_{Vmax}= -$13 days to $+$55 days. As discussed in detail in \S~\ref{id_sec}, while He lines appear to be present in its spectra and its nebular spectra show line profiles similar to other SESN, its photospheric evolution is unlike that of any other SN Ib. Thus, we designate it as SN Ib-pec. While \citet{roy13} presented their five spectra of this SN, our spectral series starts 10 days earlier than the one in \citet{roy13}. 

\vskip0.5cm SN~2008D: SN2008D and its X-ray outburst 080109 have been well studied (e.g., \citealt{soderberg08,mazzali08,modjaz09,malesani09,maund09,tanaka09}). Nine out of the ten FLWO and MMT spectra, which are amongst the earliest for a SN Ib (starting at $t_{Vmax}= -$16 days), were already presented in \citet{modjaz09}, along with spectra from other institutions and telescopes, but are included here for completeness.

\vskip0.5cm SN~2008ax: SN~2008ax was a very well-observed SN IIb (e.g., \citealt{pastorello08,roming09,taubenberger11,chornock11}) in the nearby (9.6 Mpc) NGC~4490, with a potential progenitor detection \citep{crockett08_08ax}. Here we present 25 CfA spectra that start as early as 20 days before $V$-band maximum light, i.e., only 0.8 days after the estimated shock breakout on 2008 March 3.32 \citep{pastorello08,chornock11}, and up to 1.5 days earlier than the earliest published spectra so far \citep{pastorello08,chornock11}. Our late-time spectra were included in and presented by \citet{milisavljevic10}.

\vskip0.5cm SN~2008bo: While our two MMT late-time spectra of this SN IIb were presented and analyzed in \citet{milisavljevic10}, we present here our additional 10 early-time spectra, including a number of pre-maximum spectra.

\vskip0.5cm SN~2009jf: While \citet{valenti11} and \citet{sahu09} presented very early-time spectra for this nearby SN Ib, here we present 14 new CfA spectra ranging from $t_{Vmax}= -$7 days to $+$393 days.

%%%%%%%%%%%%%%%%%%%%%%%%%%%%%%%%%%%%%%%%%%%

\section{Conclusions}\label{conclusions_sec}

While the importance of studying stripped SN and of understanding their connection to GRBs has been long acknowledged, their study has been mostly confined to individual SN. Here we present the largest spectroscopic data set of stripped Core-collapse SN to date, which comprises \ntotalspec\ densely time-sampled and homogeneous optical low-dispersion spectra of \nsntot\ SN IIb, Ib, Ic and Ic-bl, obtained through the CfA Supernova Program between 1994 and 2009. For comparison, the hitherto largest publication of early-time spectra of Stripped-Envelope Core-collapse SN by \citet{matheson01} consisted of 84 spectra of 29 SN, while the SUSPECT\footnote{http://bruford.nhn.ou.edu/~suspect/} database contains 292 spectra of 37 SESN  (including a number of SN from our sample, as well as those from \citealt{matheson01}). The recently established WISeREP database \citep{yaron12}\footnote{http://www.weizmann.ac.il/astrophysics/wiserep/} incorporates some of the SUSPECT database and contains additional spectra, though it is harder to estimate their total number of SESN spectra due to instances of double-counting. Our sample spans a large range in relatively low redshift ($<z>=$\snzmean $\pm$ \snzstdev ) and
includes spectra as early as $-$30 days before and as late as $+$611 days after $V$-band maximum (for SN~1993J up to 1000's of days) . Most of the SN in our sample have spectra before maximum light (36 out of the \nsnvmax\ SN with maximum light measured) and the mean number of spectra per SN is 9 for all SN and 12 for SN with measured maximum light. 
While not representative of SN rates, our sample consists of \nsnIIb\ SN IIb, \nsnIb\ SN Ib (where we have included SN Ib-pec 2007uy and 2009er), \nsnIc\ SN Ic, \nsnIcbl\ SN Ic-bl, and  \nsnnotsure\ SN for which either we could not determine an unique type based on our spectra (SN 1995bb, 2007iq) or which belong to an interacting SN type, probably without much hydrogen (SN 2005la, 2006jc).
The \ntotalspecfast\ spectra of stripped SN taken with the FAST
spectrograph at the FLWO 1.5m telescope were reduced in a consistent manner making this a crucial sample for detailed spectroscopic studies of  SESN and their connection to SN-GRBs (see our companion paper, M. Modjaz et al. in prep), as well as for comparisons to theoretical models and radiative transfer codes (e.g, \citealt{yoon10,dessart11,hachinger12,dessart12}) and for inclusions in SN identification schemes, both spectroscopic (e.g., \citealt{howell05,blondin07,harutyunyan08}) and photometric ones that need to include the spectrally derived k-corrections (e.g., \citealt{bernstein11}).

\acknowledgements

We are grateful for the efforts of the observers at the 1.5m FLWO, 6.5m MMT and Magellan telescopes  (listed in Table~\ref{snobs_table} in the Appendix), who obtained some of the data presented here, as well as the staffs of the F. L. Whipple, MMT and Las Campanas Observatories who assisted with and supported the CfA SN long-term projects. We thank Dan Green at the IAU CBAT for the  dissemination of SN discoveries and information, and the passionate supernova searchers, both professionals and amateurs. A specific person to whom MM is deeply grateful is Weidong Li, who was instrumental to the prolific nearby Lick Observatory SN Search over the last 14 years, as well as a dedicated and cheerful SN astronomer. We would also like to thank Alex Filippenko, Ryan Foley, Nathan Smith, Craig Wheeler, Rob Fesen, Avishay Gal-Yam, Claes Fransson, Luc Dessart, Shri Kulkarni and Alicia Soderberg for discussions. Thanks to Max Stritizinger on behalf of the Carnegie Supernova Project for providing us with the dates of maximum for SN 2004gt and 2004gv, to G. Leloudas for providing spectra of SN 2007C and 2007kj before publication, and to David Fierroz for helping to collate SN offsets.

M.M. acknowledges support from Hubble Fellowship grant HST-HF-51277.01-A, awarded by STScI, which is operated by AURA under NASA contract NAS5-26555, and from the NYU ADVANCE Women-in-Science Travel Grant funded by the NSF
ADVANCE-PAID award Number HRD-0820202, during which part of this work was performed.
Supernova research at Harvard University, including the CfA Supernova Archive, has been supported in part by
the National Science Foundation grants AST06-06772, AST09-07903 and AST-1211196. RPK is grateful to the John Simon Guggenheim Foundation and was supported by the NSF grants PHY99-07949 and NSF PHY11-25915 to the Kavli Institute for Theoretical Physics. SWJ is supported at Rutgers University in part by NSF CAREER award AST-084715 and FBB is supported by a James Arthur
fellowship at the CCPP NYU.

Furthermore, this research has made use of NASA's Astrophysics Data
System Bibliographic Services (ADS), the HyperLEDA database and the
NASA/IPAC Extragalactic Database (NED) which is operated by the Jet
Propulsion Laboratory, California Institute of Technology, under
contract with the National Aeronautics and Space Administration.  

This paper is partly based on observations obtained at the Gemini Observatory, which is
operated by the Association of Universities for Research in Astronomy,
Inc., under a cooperative agreement with the NSF on behalf of the
Gemini partnership: the National Science Foundation (United States),
the Particle Physics and Astronomy Research Council (United Kingdom),
the National Research Council (Canada), CONICYT (Chile), the
Australian Research Council (Australia), CNPq (Brazil) and CONICET
(Argentina).

 %{\it Facilities:}  \facility{MMT (Blue Channel spectrograph)}, \facility{FLWO:1.5m (FAST)}}

%%%%%%%%%%%%%%%%%%%%%%%%%%%%%%%%%%%%%%%%%%%
\clearpage
%%%%%%%%%%%%%%%% BIBLIOGRAPHY  %%%%%%%%%%%%%%%%%%%%%%%% 
\bibliographystyle{apj}
%%%%\bibliography{refs}
%\bibliography{/Users/maryammodjaz/Dropbox/refs}

%\begin{figure}
%\hskip-1.cm 
%\begin{minipage}[t]{8cm}
%\begin{center}
%\includegraphics[width=8cm]{ch4/spec/mont.sn2004ao.eps}
%\caption[Short caption for figure 1]{\label{labelFig1} Long caption figure 1}
%\end{center}
%\end{minipage}
%\hfill
%\begin{minipage}[t]{8cm}
%\begin{center}
%\includegraphics[width=8cm]{ch4/spec/mont.sn2005ek.eps}
%\caption[Short caption for figure 2]{\label{labelFig2} Long caption figure 2.}
%\end{center}
%\end{minipage}
%\end{figure}
%%%%%%%%%%%%%%%%%%%%%

%%%%%%%%%%%%%%%%%%%%%%%%%%%%%%%%%%%%%%%%%%%%%%%%%%%%%%%%%%%%%%%%%%
%%
%%   LONG Tables
%%
%%%%%%%%%%%%%%%%%%%%%%%%%%%%%%%%%%%%%%%%%%%%%%%%%%%%%%%%%%%%%%%%%%

\appendix

\section{ADDITIONAL TABLES}\label{app_sec}

\clearpage

\setcounter{table}{0}  %resets to 0 for table
\numberwithin{table}{section}

\clearpage

%\LongTables  -> for my emulateapj
\begin{turnpage}
%\begin{landscape} 
\begin{deluxetable}{lcrccccrrcccccc}
\tabletypesize{\scriptsize}
\tablewidth{550pt}
\tablecaption{Journal of Observations\label{specobs_table}}
\tablehead{
\colhead{UT Date\tablenotemark{a}} &
\colhead{HJD\tablenotemark{b}} &
\colhead{Phase\tablenotemark{c}} &
\colhead{Tel./Instr.\tablenotemark{d}} &
\colhead{Range\tablenotemark{e}} &
\colhead{Disp.\tablenotemark{f}} &
\colhead{Res.\tablenotemark{g}} &
\colhead{P.A.\tablenotemark{h}} &
\colhead{$|\Delta\Phi|$\tablenotemark{i}} &
\colhead{Air.\tablenotemark{j}} &
\colhead{Flux Std.\tablenotemark{k}} &
\colhead{See.\tablenotemark{l}} &
\colhead{Slit\tablenotemark{m}} &
\colhead{Exp.\tablenotemark{n}} &
\colhead{Observer(s)\tablenotemark{o}} \\
\colhead{} &
\colhead{} &
\colhead{(d)} &
\colhead{} &
\colhead{(\AA)} &
\colhead{(\AA/pix)} &
\colhead{(\AA)} &
\colhead{(\degr)} &
\colhead{(\degr)} &
\colhead{} &
\colhead{} &
\colhead{(\arcsec)} &
\colhead{(\arcsec)} &
\colhead{(s)} &
\colhead{}
}
\startdata
\multicolumn{15}{c}{\bf SN 1994I} \\
1994-04-03.49 & 2449445.99 &  $-$5.9 &     MMT+Blue Channel &  3483-7697 & 3.51 &   7-8 &    \nodata &       81.2 & 1.38 &      \nodata & \nodata & \nodata &                  120 &              GB \\
1994-04-03.39 & 2449445.89 &  $-$6.0 &       FLWO 1.5m+FAST &  3705-7635 & 1.47 &   6-7 &       90.0 &       50.8 & 1.07 &      F67/F56 &       2 &     3.0 &                  900 &             JPe \\
1994-04-05.32 & 2449447.82 &  $-$4.1 &       FLWO 1.5m+FAST &  3704-7635 & 1.47 &   6-7 &       90.0 &       60.9 & 1.05 &      F67/F56 &     2-3 &     3.0 &         3$\times$900 &             PBe \\
1994-04-06.39 & 2449448.89 &  $-$3.0 &       FLWO 1.5m+FAST &  3707-7636 & 1.47 &   6-7 &       90.0 &       45.6 & 1.08 &      F67/F56 &     2-3 &     3.0 &         2$\times$900 &             PBe \\
1994-04-07.38 & 2449449.88 &  $-$2.1 &       FLWO 1.5m+FAST &  3707-7637 & 1.47 &   6-7 &       90.0 &       48.2 & 1.08 &      F67/F56 &     1-2 &     3.0 &         2$\times$900 &             PBe \\
1994-04-08.33 & 2449450.83 &  $-$1.1 &       FLWO 1.5m+FAST &  3706-7636 & 1.47 &   6-7 &       90.0 &       83.0 & 1.04 &      F67/F56 &     1-2 &     3.0 &                  900 &             JPe \\
1994-04-10.27 & 2449452.77 &    +0.8 &       FLWO 1.5m+FAST &  3705-7635 & 1.47 &   6-7 &       90.0 &       35.3 & 1.07 &      F67/F56 &     1-2 &     3.0 &                  900 &             JPe \\
1994-04-110.5 & 2449454.00 &    +2.1 &     MMT+Blue Channel &  3209-8760 & 1.91 &   7-8 &    \nodata &       82.9 & 1.62 &      \nodata & \nodata & \nodata &                  300 &      JHuc, CHel \\
1994-04-11.25 & 2449453.75 &    +1.8 &       FLWO 1.5m+FAST &  3705-7635 & 1.47 &   6-7 &       90.0 &       26.8 & 1.09 &      F67/F56 &     1-2 &     3.0 &         2$\times$900 &             PBe \\
1994-04-12.44 & 2449454.94 &    +3.0 &       FLWO 1.5m+FAST &  3708-7638 & 1.47 &   6-7 &       90.0 &       11.9 & 1.26 &      F67/F56 &     1-2 &     3.0 &         2$\times$900 &             PBe \\
1994-04-15.37 & 2449457.87 &    +5.9 &       FLWO 1.5m+FAST &  4844-6839 & 0.75 &   3-4 &      170.0 &       35.8 & 1.08 &         HZ44 &     1-2 &     3.0 &                  900 & RJ, Fabricant,  \\
1994-04-16.35 & 2449458.85 &    +6.9 &       FLWO 1.5m+FAST &  4843-6839 & 0.75 &   3-4 &   $-$177.0 &       71.0 & 1.04 &         HZ44 &     1-2 &     3.0 &                  900 &   RJ, Fabricant \\
1994-04-30.35 & 2449472.85 &   +20.9 &       FLWO 1.5m+FAST &  3707-7631 & 1.47 &   6-7 &      110.0 &        8.5 & 1.14 &      F67/F56 &     1-2 &     3.0 &         2$\times$900 &             PBe \\
1994-05-01.39 & 2449473.89 &   +21.9 &       FLWO 1.5m+FAST &  3708-7632 & 1.47 &   6-7 &      110.0 &        3.0 & 1.25 &      F67/F56 &     1-2 &     3.0 &                 1800 &             PBe \\
1994-05-02.37 & 2449474.87 &   +22.9 &       FLWO 1.5m+FAST &  3708-7631 & 1.47 &   6-7 &      110.0 &        0.2 & 1.22 &      F67/F56 &     1-2 &     3.0 &                 1800 &             PBe \\
1994-05-03.28 & 2449475.78 &   +23.8 &       FLWO 1.5m+FAST &  3706-7630 & 1.47 &   6-7 &      110.0 &       68.6 & 1.05 &      F67/F56 &     1-2 &     3.0 &                 1800 &             JPe \\
1994-05-05.20 & 2449477.70 &   +25.7 &       FLWO 1.5m+FAST &  3704-7628 & 1.47 &   6-7 &      110.0 &       45.8 & 1.08 &      F67/F56 &     1-2 &     3.0 &                 1800 &             JPe \\
1994-05-07.31 & 2449479.81 &   +27.8 &       FLWO 1.5m+FAST &  3707-7631 & 1.47 &   6-7 &      110.0 &       26.0 & 1.10 &      F67/F56 &     1-2 &     3.0 &                 1800 &             PBe \\
1994-05-08.26 & 2449480.76 &   +28.8 &       FLWO 1.5m+FAST &  3708-7632 & 1.47 &   6-7 &      110.0 &       88.0 & 1.24 &      F67/F56 &     1-2 &     3.0 &                 1200 &             PBe \\
1994-05-09.26 & 2449481.76 &   +29.8 &       FLWO 1.5m+FAST &  3707-7631 & 1.47 &   6-7 &      110.0 &       70.2 & 1.05 &      F67/F56 &     1-2 &     3.0 &                 1800 &             JPe \\
1994-05-13.30 & 2449485.80 &   +33.8 &       FLWO 1.5m+FAST &  3708-7632 & 1.47 &   6-7 &      110.0 &       15.1 & 1.11 &      F67/F56 &     1-2 &     3.0 &                  900 &             PBe \\
1994-05-15.22 & 2449487.72 &   +35.7 &       FLWO 1.5m+FAST &  3720-7643 & 1.47 &   6-7 &      110.0 &       86.1 & 1.04 &      F67/F56 &     2-3 &     3.0 &                  900 &             SMu \\
1994-05-16.22 & 2449488.72 &   +36.7 &       FLWO 1.5m+FAST &  3710-7634 & 1.47 &   6-7 &      110.0 &       85.7 & 1.04 &      F67/F56 &     2-3 &     3.0 &         2$\times$900 &             SMu \\
1994-05-17.25 & 2449489.75 &   +37.8 &       FLWO 1.5m+FAST &  3744-7668 & 1.47 &   6-7 &      110.0 &       44.3 & 1.05 &      F67/F56 &     2-3 &     3.0 &                  900 &             SMu \\
1994-05-19.19 & 2449491.69 &   +39.7 &       FLWO 1.5m+FAST &  3747-7670 & 1.47 &   6-7 &      110.0 &       72.2 & 1.05 &      F67/F56 &     2-3 &     3.0 &                 1200 &             JPe \\
1994-06-04.27 & 2449507.77 &   +55.7 &       FLWO 1.5m+FAST &  3799-7723 & 1.47 &   6-7 &      110.0 &        0.1 & 1.19 &      F67/F56 &     2-3 &     3.0 &                  900 &             SMu \\
\enddata
\tablecomments{This table is available in its entirety in a machine-readable form in the online journal. A portion is shown here for guidance regarding its form and content.}

\tablenotetext{a}{UT at midpoint of observation(s).}
\tablenotetext{b}{Heliocentric Julian date at midpoint of observation(s).}
\tablenotetext{c}{Rest-frame age of spectrum in days relative to $V$-band maximum. For \nsnvnomax\ SNe with no reliable estimate for the time of maximum, we indicate the rest-frame days relative to the first spectrum preceded by an ``@'' symbol}
\tablenotetext{d}{Telescope and instrument used for this spectrum: FAST = FLWO 1.5m+FAST, IMACS = Magellan Baade+IMACS, LDSS2 = Magellan Clay+LDSS2, LDSS3 = Magellan Clay+LDSS3, MMTblue = MMT+Blue Channel, MMTred = MMT+Red Channel.}
\tablenotetext{e}{Observed wavelength range of spectrum.}
\tablenotetext{f}{Spectral dispersion in \AA\ per pixel.}
\tablenotetext{g}{Approximate FWHM spectral resolution in \AA.}
\tablenotetext{h}{Observed position angle during the observation(s).}
\tablenotetext{i}{Absolute difference between the observed position angle and the average parallactic angle over the course of the observation(s).}
\tablenotetext{j}{Airmass of the observation.}
\tablenotetext{k}{Standard stars: 
BD17~=~BD+17$^{\circ}$4708,
BD26~=~BD+26$^{\circ}$2606,
BD28~=~BD+28$^{\circ}$4211,
BD33~=~BD+33$^{\circ}$2642,
CD32~=~CD-32~9927,
EG131~=~EG~131,
EG274~=~EG~274,
F15~=~Feige~15,
F25~=~Feige~25,
F34~=~Feige~34,
F56~=~Feige~56,
F66~=~Feige~66,
F67~=~Feige~67,
F110 ~=~Feige~110,
G191~=~G191B2B,
H102~=~Hiltner~102,
H600~=~Hiltner~600,
HD19~=~HD~192281,
HD21~=~HD~217086,
HD84~=~HD~84937,
HZ44~=~HZ~44,
HZ14~=~HZ~14,
L3218~=~LTT~3218,
L3864~=~LTT~3864,
L4816~=~LTT~4816,
vMa2~=~van Maanen 2.
}
\tablenotetext{l}{Seeing is based upon estimates by the observers.}
\tablenotetext{m}{Spectroscopic slit width.}
\tablenotetext{n}{Exposure time. Separate exposures are indicated.}
\tablenotetext{o}{Observers:
EA~=~E.~Adams,
VA~=~V.~Antoniou,
HA~=~H.~Arce,
JA~=~J.~P.~Anderson,
ZB~=~Z.~Balog,
PBa~=~P.~Barmby,
EB~=~E.~Barton,
JB~=~J.~Battat,
PBe~=~P.~Berlind,
GB~=~G.~Bernstein,
WBl~=~W.~P.~Blair,
SB~=~S.~Blondin,
AB~=~A.~E.~Bragg,
CB~=~C.~Brice\~no,
WBr~=~W.~Brown,
NC~=~N.~Caldwell,
MC~=~M.~L.~Calkins,
BC~=~B.~J.~Carter,
PC~=~P.~Challis,
JC~=~J.~R.~Cho,
CC~=~C.~Clemens,
ACo~=~A.~Cody,
ACr~=~A.~Crook,
TC~=~T.~Currie,
RC~=~R.~M.~Cutri,
KD~=~K.~Dendy,
AD~=~A.~Diamond-Stanic,
JDon~=~J.~L.~Donley,
JDow~=~J.~J.~Downes,
KE~=~K.~Eriksen,
GE~=~G.~Esquerdo,
DF~=~D.~Fabricant,
EF~=~E.~E.~Falco,
RF~=~R.~Fesen,
JF~=~J.~Foster,
JGa~=~J.~Gallagher,
AG~=~A.~Garg,
PG~=~P.~M.~Garnavich,
IG~=~I.~Ginsburg,
JGr~=~J.~Graves,
NG~=~N.~Grogin,
TG~=~T.~Groner,
VH~=~V.~Hradecky,
HH~=~H.~Hao,
CHei~=~C.~Heinke,
CHel~=~C.~Heller,
JHe~=~J.~Hernandez,
MH~=~M.~Hicken,
CHi~=~C.~Hill,
JHuc~=~J.~P.~Huchra,
JHug~=~J.~P.~Hughes,
CHu~=~C.~Hutcheson,
RH~=~R.~Hutchins,
RJ~=~R.~Jansen,
SJ~=~S.~Jha,
SKa~=~S.~J.~Kannapan,
SKe~=~S.~Kenyon,
RK~=~R.~P.~Kirshner,
DK~=~D.~M.~Koranyi,
JK~=~J.~Kuraszkiewicz,
HL~=~H.~Landt,
TL~=~T.~Lappin,
NL~=~N.~Lepore,
LM~=~L.~Macri,
JM~=~J.~A.~Mader,
AM~=~ A.~Mahdavi,
EM~=~E.~Mamajek,
SMa~=~S.~A.~Mao,
NM~=~N.~Martimbeau,
TM~=~T.~Matheson,
MM~=~M.~Modjaz,
FM~=~F.~Munshi,
SMu~=~S.~Muscarella,
GN~=~G.~Narayan,
PN~=~P.~Nutzman,
CP~=~C.~A.~Pantoja,
BP~=~B.~M.~Patten,
KP~=~K.~Penev,
JPe~=~J.~Peters,
WP~=~W.~Peters,
MP~=~M.~Phelps,
JPi~=~J.~Pi\~neda,
AR~=~A.~G.~Riess,
KR~=~K.~Rines,
MS~=~M.~Schr\"odter,
JS~=~J.~D.~Silverman,
IS~=~I.~Song,
ST~=~S.~Tokarz,
MT~=~M.~Torres,
CT~=~C.~Tremonti,
AV~=~A.~Vaz,
LW~=~L.~Wells,
MW~=~M.~Westover.
}
\label{snobs_table}

\end{deluxetable}
%\end{landscape}

\end{turnpage}

\clearpage
%%%% FLUX CALIBRATION DETAILS
\section{Relative Flux Calibration}\label{flux_sec}

Here we check the accuracy of the relative flux calibration of our stripped SN spectra across the spectrum, employing a similar, but improved, procedure as for the CfA SN Ia data in \citet{matheson08} and in \citet{blondin12}. We compared the $B-V$ colors calculated from our stripped SN spectra with those based on the corresponding calibrated CfA light curves of the same stripped SN in the Bessel standard star system \citep{bianco14}. We show the results in Figure~\ref{color_fig}. For some spectra that are heavily contaminated by HII regions, we first removed the HII region emission lines, and then computed spectra-based $(B-V)$ colors by multiplying the spectra with Bessel $B$ and $V$ filter functions. Furthermore, we only used spectra that were taken at parallactic angle in order to test for effects on spectrophotometry other than the known differential light loss produced by atmospheric dispersion. Moreover, we computed photometry-based colors only if they were available within one day of the spectra. Lastly, we propagated the photometry uncertainties into the color calculations, a new feature compared to similar plots of \citet{matheson08} and \citet{blondin12}, and plot them on Figure \ref{color_fig}. For our stripped SN sample, which consists of 116 data points, we find that the scatter around zero difference between spectra-based and photometry-based colors is $\sim$2 times smaller for spectra at phases less than 20 days ($\sim$0.15 mag) than for phases larger than 20 days ($\sim$0.37 mag). This factor of two is consistent with the conclusions based on SN Ia data presented in \citet{matheson08} and \citet{blondin12}. As mentioned in those works, the likely reasons are twofold: 1) over time, the spectra become dominated by strong emission lines, such that any deviation of the actual filter response function from the assumed one, with which we convolved our spectra, introduces errors, and 2.) the SN get fainter over time, which makes host galaxy contamination more significant. We find no significant difference for the color comparison between SN Ib and SN Ic. 

For comparison, we overplot the corresponding colors for the SNe Ia data. For consistency, we down-loaded the SN Ia data from the CfA SN Archive (using spectra from \citealt{blondin12} and photometry from \citealt{riess99,jha06,hicken09}) and re-analyzed them by applying the same procedure as used for the stripped SN data (see above), yielding 348 data points for SN Ia. In order to quantify any systematic deviation from the diagonal, i.e. the one-to-one correspondence of both colors, we performed a simple weighted least-squares-fit to each SN sample (for all phases, over the same color-range as for stripped SN), and computed the scatter around the fitted line. 
While both SN samples appear to show somewhat bluer colors from spectra than from photometry for a few SN, the formal weighted fits yield only slight deviations from a diagonal (see fitted lines, slope for SN Ia: 1.037$\pm$0.019, slope for stripped SN: 1.033 $\pm$ 0.038).  SN Ia have less scatter around the fitted line (from our analysis: 0.12 mag, including all phases) than stripped SN (0.24 mag, including all phases). However, these values may not necessarily imply that the relative flux calibration for the stripped SN spectra is twice worse than for SN Ia spectra, given that the \emph{photometry data} of stripped SN is as well worse than for SN Ia, as evidenced by the larger error bars for the stripped SN data than for SN Ia. Furthermore,  the fact that the stripped SN sample has a factor of two worse scatter than the SN Ia sample could partly be due to the fact that there are more older stripped SN than older SN Ia in this comparison: 34\% and 27\%, respectively, of the data points for the stripped SN and for the SN Ia, respectively, are at phases later than 20 days.

\setcounter{figure}{0}  %resets to 0 for figure in appendix
\numberwithin{figure}{section}

%%%%%%%%%
\begin{figure}[!ht]
%\hspace-0.5
\includegraphics[scale=0.45,angle=0]{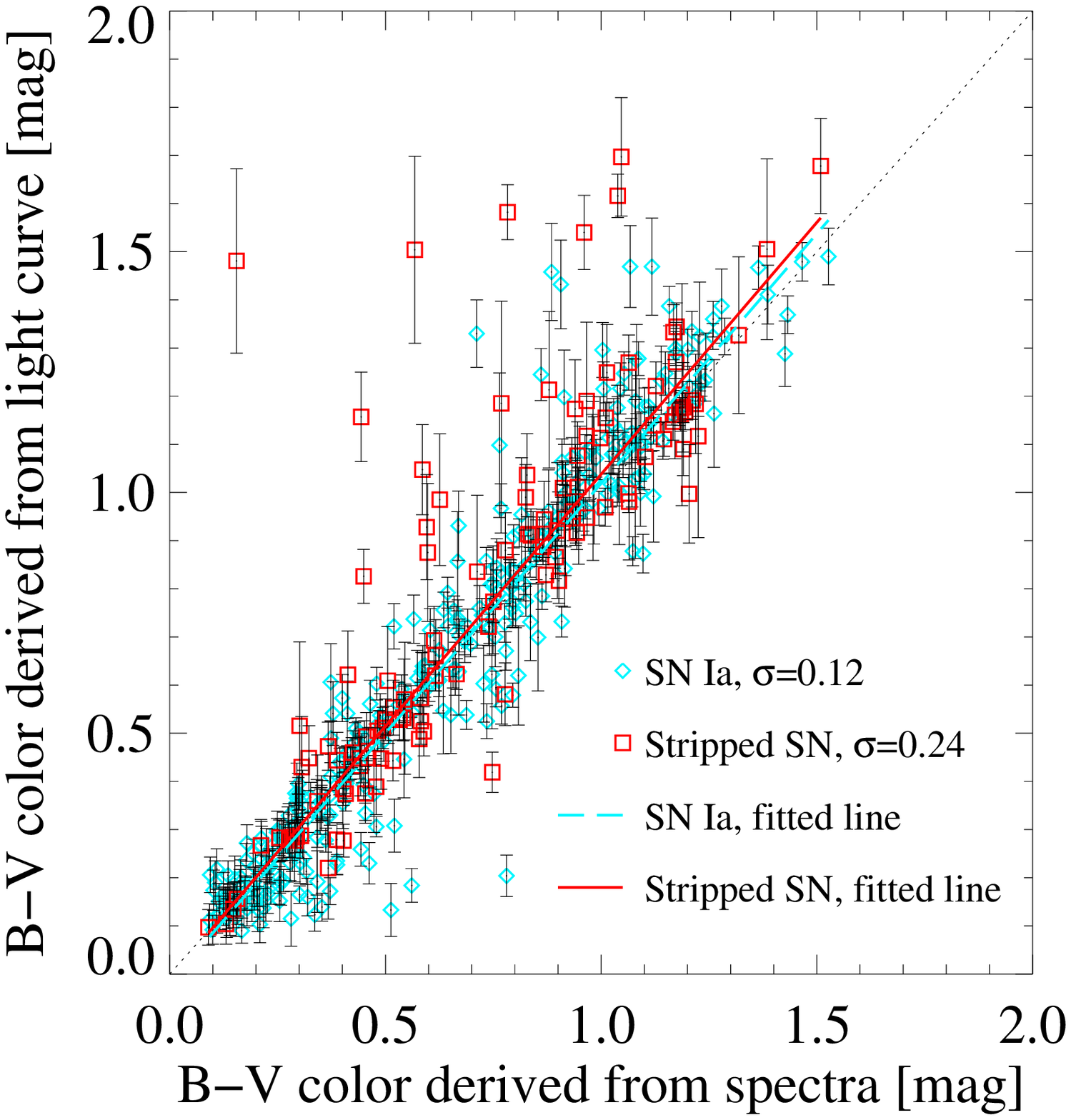} 
\singlespace \caption{Comparison of the colors of stripped SN in our sample derived from spectroscopy (this work) and from photometry \citep{bianco14} over all phases, in order to evaluate the relative flux calibration of the CfA spectra. Note that the axes are inverted compared to corresponding figures in \citet{matheson08} and \citet{blondin12} for the CfA data of SN Ia. The full stripped SN sample has a scatter (0.24 mag) that is a factor of $\sim$2 higher than for the full SN Ia sample (0.12 mag, using spectra from \citealt{blondin12} and photometry from \citealt{riess99,jha06,hicken09}, re-analyzed in the same way as the stripped SN data).}
\label{color_fig}
\end{figure}
%%%%%%%%%%

\clearpage

%%%%%% TABLE: SN SAMPLE 
%\input{ch4/SNSample_table.tex}
%%%\label{snsample_table}
%%%%%%%%%%%%%%%

\begin{deluxetable}{llllllllc}
\tabletypesize{\scriptsize}
\singlespace
\tablecaption{Stripped SN Sample: SN and Host Galaxy Information}
\tablehead{\colhead{SN Name} & 
\colhead{SN Type\tablenotemark{a}} & 
\colhead{Disc Ref} & 
\colhead{Disc Type\tablenotemark{b}} & 
\colhead{Host Galaxy} & 
\multicolumn{2}{c}{SN Offsets\tablenotemark{c}} & 
\colhead{$cz_{\rm{helio}}$} & 
\colhead{JD of SN $V$-max (Ref\tablenotemark{d})} \\ 
\colhead{} &
\colhead{} &
\colhead{} &
\colhead{} &
\colhead{} &
\colhead{[\arcsec]} &
\colhead{[\arcsec]} &
\colhead{[\kms]} &
\colhead{} 
} 
\startdata
   1993J &        IIb & IAUC        5731 &     T &                 M81 &    45W &   160S &     -140 &  2449095.0 (     R94) \\
   1994I &         Ic & IAUC        5961 &     T &                 M51 &    14E &    12S &      461 &  2449451.4 (     R96) \\
   1995F &         Ib & IAUC        6165 &     T &             NGC2726 &     2E &     1S &     1440 &     \ldots (  \ldots) \\
  1995bb &       Ib/c & IAUC        6271 & Non-T &               Anonymous & $<$1.5 & $<$1.5 &     1610 &     \ldots (  \ldots) \\
  1996cb &        IIb & IAUC        6524 &     T &             NGC3510 &  20.9W &  65.7N &      800 &  2450454.4 (     Q99) \\
   1997X &         Ib & IAUC        6552 &     T &             NGC4691 &   7.2E &   0.3N &     1072 &     \ldots (  \ldots) \\
  1997dq &         Ic & IAUC        6770 &     T &             NGC3810 &    43W &    29N &      993 & $<$2450754 (    IAUC) \\
  1997ef &      Ic-bl & IAUC        6778 &     T &             UGC4107 &    10E &    20S &     3400 &  2450793.8 (     I00) \\
  1998dt &         Ib & IAUC        7006 &     T &              NGC945 &  23.1W &  39.5S &     4580 &  2451068.5\tablenotemark{e} (     M01) \\
  1998fa &        IIb & IAUC        7073 &     T &             UGC3513 &   4.0W &   3.8N &     7460 &  2451180.5\tablenotemark{e} (     M01) \\
   2000H &        IIb & IAUC        7361 &     T &               IC454 &  19.6E &   0.2S &     3894 &  2451585.5 (     B02) \\
  2001ai &         Ib & IAUC        7605 &     T &             NGC5278 &   2.0W &   9.0S &     7753 &     \ldots (  \ldots) \\
  2001ej &         Ib & IAUC        7719 &     T &             UGC3829 &     4W &     7N &     4135 & $<$2452188 (     CfA) \\
  2001gd &        IIb & IAUC        7761 &     T &             NGC5033 &    52W &   161N &     1024 & $<$2452267 (     CfA) \\
  2002ap &      Ic-bl & IAUC        7810 &     T &                 M74 &   258W &   108S &      659 &  2452313.4 (     F03) \\
  2002ji &         Ib & IAUC        8025 &     T &             NGC3655 &  22.4W &  14.0S &     1624 &     \ldots (  \ldots) \\
  2003jd &      Ic-bl & IAUC        8232 &     T &        MCG-01-59-21 &   8.3E &   7.7S &     5635 &  2452943.3 (     V08, CfA) \\
  2004aw &         Ic & IAUC        8310 &     T &             NGC3997 &  27.7E &  19.8S &     4721 &  2453091.5 (     T06, CfA ) \\
  2004ao &         Ib & IAUC        8299 &     T &            UGC10862 &   6.3E &  23.8S &     1691 & $<$2453079 (     CfA) \\
  2004dk &         Ib & IAUC        8377 &     T &             NGC6118 &   4.4E &  43.5N &     1573 &  2453238.0 (     D11) \\
  2004dn &   Ic & IAUC        8381 &     T &             UGC2069 &   0.5W &  26.1S &     3777 &  2453230.5 (     D11) \\
  2004eu &         Ic\tablenotemark{f}  & IAUC        8446 & Non-T &         MCG+07-5-39 &   1.5E &   2.7N &     6486 &     \ldots (  \ldots) \\
  2004fe &         Ic & IAUC        8417 &     T &              NGC132 &   2.1W &   7.9S &     5361 &  2453319.3 (    CfA) \\
  2004ff &         IIb & IAUC        8425 &     T &          ESO552-G40 &   8.7E &  12.3S &     6664 &  2453316.0\tablenotemark{e} (     D11) \\
  2004ge &         Ic & IAUC        8425 &     T &             UGC3555 &  11.2W &   2.8S &     4644 &  2453341.5\tablenotemark{e} (     D11) \\
  2004gk &         Ic & IAUC        8443 &     T &              IC3311 &   6.2E &   1.4S &     -122 & $<$2453338 (     CfA) \\
  2004gq &         Ib & IAUC        8452 &     T &             NGC1832 &  22.3E &  22.4N &     1859 &  2453361.0 (     D11) \\
  2004gt &         Ic & IAUC        8454 &     T &             NGC4038 &    34W &    10S &     1642 &  2453362.0 (     S) \\
  2004gv &         Ib & IAUC        8454 &     T &              NGC856 &  13.8W &   4.0S &     5974 &  2453366.2 (     S) \\
   2005U &        IIb & IAUC        8473 &     T  &              Arp299 &   3.7W &   4.9S &     3021 &     \ldots (  \ldots) \\
  2005ar &         Ib & IAUC        8493 &     T &         CGCG011-033 &   4.6E &   5.6N &     7629 &     \ldots (  \ldots) \\
  2005az &         Ic & IAUC        8503 & Non-T &             NGC4961 &     8W &   5.5N &     2622 &  2453473.9 (     CfA) \\
  2005bf &         Ib & IAUC        8507 &     T &        MCG+00-27-00 &  11.7E &  32.6S &     5430 &  2453499.0\tablenotemark{g} (     CfA) \\
  2005da &      Ic-bl & IAUC        8570 & Non-T &            UGC11301 & 108.4W &  34.8N &     4496 &     \ldots (  \ldots) \\
  2005hg &         Ib & IAUC        8623 &     T &             UGC1394 &   3.7W &  19.6S &     6388 &  2453684.4 (     CfA) \\
  2005kf &         Ic & IAUC        8630 & Non-T & J074726.40+265532.4 &   0.8E &    0.6S &     4522 &  $<$2453699 (  CfA) \\
  2005kl &         Ic & CBET         300 &     T &             NGC4369 &     6W &     4N &     1045 & 2453703.7 (     CfA) \\
  2005la &     Ib-n/IIb-n & IAUC        8639 & Non-T &         KUG1249+278 &     6W &     6S &     5570 &  2453694.5\tablenotemark{e} (    P08a) \\
  2005mf &         Ic & IAUC        8648 &     T &            UGC04798 &   5.9W &  13.3N &     8023 &   2453734.56 (     CfA) \\
  2005nb &      Ic-bl & CBET         357 & Non-T &             UGC7230 &   1.5W &    5.N &     7127 & $<$2453747.9 (     CfA) \\
   2006T &        IIb & IAUC        8666 &     T &             NGC3054 &    22E &    21S &     2650 &  2453782.1 (     CfA) \\
  2006aj &      Ic-bl & IAUC        8674 & Non-T &        J032139.68+165201.7 & $<$0.2 & $<$0.2 &    10052 &  2453794.7 (     M06) \\
  2006ck &         Ic & IAUC        8713 &     T &            UGC08238 &  10.2W &   3.0S &     7312 & $<$2453879 (     CfA) \\
  2006el &        IIb & IAUC        8741 &     T &            UGC12188 &  12.9E &  18.2S &     5150 &  2453984.7 (     CfA, D11) \\
  2006ep &         Ib & IAUC        8744 &     T &              NGC214 &    43W &  11.3S &     4505 &  2453988.7 (     CfA) \\
  2006fo &         Ib & IAUC        8750 & Non-T &            UGC02019 &   6.0W &   0.6N &     6214 & $<$2454006 (     CfA, D11) \\
  2006jc &       Ib-n & IAUC        8762 &     T &            UGC04904 &    11W &     7S &     1670 &  2454020.0 (     F07) \\
  2006lc &         Ib & CBET         693 & Non-T &             NGC7364 &   1.4E &  10.0S &     4863 & $<$2454045 (     CfA) \\
  2006ld &         Ib & IAUC        8766 & Non-T &              UGC348 &     1E &    18S &     4179 & $<$2454044 (     CfA) \\
  2006lv &         Ib & IAUC        8771 &     T &             UGC6517 &    10E &    12N &     2570 & $<$2454043 (     CfA) \\
   2007C &         Ib & CBET         798 &     T &             NGC4981 &     9E &    22S &     1767 &   2454115.9 (     CfA) \\
   2007D &      Ic-bl & IAUC        8794 &     T &             UGC2653 &   4.9E &   2.8S &     6944 &  2454123.3 (     CfA) \\
   2007I &      Ic-bl & IAUC        8798 & Non-T & J115913.13-013616.1 &   0.8E &   0.8S &     6487 & $<$2454117 (     CfA) \\
  2007ag &         Ib & CBET         868 &     T &             UGC5392 &   4.0E &  15.5N &     6017 &  2454170.3 (     CfA) \\
  2007bg &      Ic-bl & IAUC        8834 & Non-T &           Anonymous & $<$0.2 & $<$0.2 &    10370 &  2454206.7\tablenotemark{e} (     Y10) \\
  2007ce &      Ic-bl & IAUC        8843 & Non-T &           J121018.03+484331.8 & $<$0.5 & $<$0.5 &    13890 & $<$2454226 (     CfA) \\
  2007cl &         Ic\tablenotemark{f}  & IAUC        8851 &     T &             NGC6479 &   3.2W &   8.2N &     6650 &  2454250.9 (     CfA) \\
  2007gr &         Ic & IAUC        8864 &     T &             NGC1058 &  24.8W &  15.8N &      518 &  2454338.5 (     H09) \\
  2007hb &         Ic & CBET        1064 &     T &              NGC819 &  39.9W &   3.9N &     6669 & $<$2454363 (     CfA) \\ %lik e07gr
  2007iq &   Ic/Ic-bl & CBET        1043 & Non-T &             UGC3416 &    19W &     5S &     4003 & $<$2454363 (     CfA) \\
  2007kj &         Ib & CBET        1092 &     T &             NGC7803 &     6W &    10S &     5366 &  2454382.5 (     CfA) \\
  2007ru &      Ic-bl & CBET        1149 &     T &            UGC12381 &   4.4E &  39.8S &     4636 &  2454440.5 (    CfA, S09) \\
  2007rz &         Ic & CBET        1158 &     T &             NGC1590 &   8.6E &   0.1N &     3897 & $<$2454454 (     CfA) \\
  2007uy &         Ib-pec & IAUC        8908 &     T &             NGC2770 &  20.6E &  15.5S &     2100 &  2454481.8 (     CfA) \\
   2008D &         Ib & CBET        1202 &     T &             NGC2770 &  38.3W &  55.6N &     2100 &   2454494.6 (    S08, M09, CfA) \\
  2008an &         Ic & CBET        1268 &     T &            UGC10936 &   4.9E &   9.8S &     8124 & $<$2454524 (     CfA) \\
  2008aq &        IIb & CBET        1271 &     T &        MCG-02-33-20 &  15.2E &  45.2S &     2389 & $<$2454527 (     CfA) \\
  2008ax &        IIb & CBET        1280 &     T &             NGC4490 &  53.1E &  25.8S &      630 &  2454549.5 (P08b;C11) \\
  2008bo &        IIb & CBET        1324 &     T &             NGC6643 &    31E &    15N &     1484 &  2454569.7 (     CfA) \\
  2008cw &        IIb & CBET        1395 & Non-T & J163238.16+412730.8 &     1E &     2N &     9724 & $<$2454622 (     CfA) \\
  2009er &         Ib-pec & CBET        1811 &     T & J153930.49+242614.8 &     8W &     9S &    10500 &  2454982.6 (     CfA) \\
  2009iz &         Ib & CBET        1947 &     T &             UGC2175 &  12.3W &  14.5N &     4257 &  2455109.4 (     CfA) \\
  2009jf &         Ib & CBET        1952 &     T &             NGC7479 &  53.8W &  36.5N &        2443 &	2455122.8 (     CfA, V11)
\enddata
\tablenotetext{a}{The SN types were re-determined by us via running SNID on all our SN spectra and are sometimes different from those listed in the IAU list of SNe (see Table~\ref{snidchange_table}). The SN Type "Ic-bl" stands for broad-lined SN Ic. SN~2006jc is a SN Ib-n ("n" for narrow) which showed narrow He lines in emission. See text  for details. }
\tablenotetext{b}{Discovery type: "T" = SN host galaxy was targeted; "Non-T" = SN host galaxy was not targeted..}
\tablenotetext{c}{Offsets from the center of host galaxy as listed in discovery IAUCs or as derived by comparing SN and host-galaxy coordinates.}
\tablenotetext{d}{References: R94 = \citet{richmond94}. R96 = \citet{richmond96}. Q99 = \citet{qui99}. I00 = \citet{iwamoto00}. M01 = \citet{matheson01}. CfA = CfA photometry presented in \citet{bianco14}, where the date of $V$-band maximum is derived by fitting Monte Carlo realizations of the V-band photometric data points near peak with a second-degree polynomial. For cases where literature data were available for the same SN, those data were also included in the Þt. B02 = \citet{branch02}. F03 = \citet{foley03}. V08 = \citet{valenti08}. T06 = \citet{taubenberger06}. D11 = \citet{drout11}. S = Stritzinger et al. (in prep). T05= \citet{tominaga05}. Note that the V-band light curve of SN 2005bf had two distinct maxima, and that we are using the first maximum as our reference. P08a = \citet{pastorello08_05la}. M06 = \citet{modjaz06}. F07 = \citet{foley07}. Y10 = \citet{young10}. H09 = \citet{hunter09}. S09 = \citet{sahu09}. S08 = \citet{soderberg08}. M09 = \citet{modjaz09}. P08b = \citet{pastorello08}. C11 = \citet{chornock11}. V11 = \citet{valenti11}.  Upper limits on date of maximum are indicated and are based on the CfA data.
}
\tablenotetext{e}{SNe with measured and listed maximum only in R-band.}
\tablenotetext{f}{We cannot exclude the potential possibility of the emergence of \He\ lines in these SN Ic, since our spectra were either taken before or well after maximum light. }
\tablenotetext{g}{SN~2005bf showed a pronounced double-peak light curve - here we are using the main, second, maximum.}

\label{snsample_table}
\end{deluxetable}

\end{document}